\documentclass[12pt]{article}
\usepackage{amsmath}
\usepackage{amssymb}
\usepackage{mathtools}
\newcommand{\be}{\begin{equation}}
\newcommand{\bea}{\begin{eqnarray}}
\newcommand{\ee}{\end{equation}}
\newcommand{\eea}{\end{eqnarray}}

\begin{document}
\topmargin -1cm \oddsidemargin=0.25cm\evensidemargin=0.25cm
\setcounter{page}0
\renewcommand{\thefootnote}{\fnsymbol{footnote}}
\begin{titlepage}
\begin{flushright}
LTH-978
\end{flushright}
\vskip .7in
\begin{center}
{\Large \bf Massive Higher Spin Fields Coupled to a Scalar: \\
Aspects of Interaction and  Causality
 } \vskip .7in {\large I.L. Buchbinder$^a$\footnote{e-mail: {\tt joseph@tspu.edu.ru}},
P. Dempster$^b$\footnote{e-mail: {\tt  paul.dempster@liv.ac.uk }} and
 M.Tsulaia$^c$\footnote{e-mail: {\tt  mirian.tsulaia@canberra.edu.au}  }}
\vskip .2in {$^a$ \it Department of Theoretical Physics, Tomsk State Pedagogical University,
Tomsk, 634061, Russia} \\
\vskip .2in {$^b$ \it
Department of Mathematical Sciences, University of Liverpool,
Liverpool, L69 7ZL, United Kingdom} \\
\vskip .2in { $^c$ \it Faculty of  Education Science Technology and Mathematics, University of
Canberra, Bruce ACT 2617, Australia}\\
\begin{abstract}

We consider in detail the most general cubic Lagrangian which
describes  an interaction between two identical higher spin fields
in a triplet formulation with a scalar field,  all fields having the
same values of the mass. After performing the gauge fixing procedure
we find that for the case of massive fields  the gauge invariance
does not guarantee the preservation of the correct number of
propagating physical degrees of freedom. In order to get the correct
number of degrees of freedom for the massive higher spin field one
should impose some additional conditions on parameters of the vertex.
Further independent constraints are provided by the causality
analysis, indicating that the requirement of causality should be
imposed in addition to the requirement of gauge invariance in order
to have a consistent propagation of massive higher spin fields.

\end{abstract}

\end{center}

\vfill

\end{titlepage}

\section{Introduction}

It is already well known that massless higher spin fields can interact consistently
on an arbitrary dimensional anti-de Sitter $AdS_{\cal D}$
background \cite{Vasiliev:2003ev}.
On the other hand,
the situation is quite different as far as interacting theories for higher spin fields
on a flat background  are concerned.
 The corresponding  cubic interaction vertices for massless fields, whose study has been initiated in
\cite{Bengtsson:1983pd}--\cite{Koh:1986vg},
are fairly well understood  now
\cite{Metsaev:2005ar}-- \cite{Henneaux:2012wg}.
The application of cubic vertices on a flat background can be twofold.
The first one is to use  them
 for a further deformation to $AdS_{\cal D}$, thus studying the
cubic interactions in the ``metric like" formalism
 \cite{Fotopoulos:2007yq}--\cite{Joung:2011ww}.\footnote{Cubic and higher--order interactions  in the ``frame--like"
formalism  have been extensively discussed in \cite{Vasilev:2011xf}--\cite{Boulanger:2013zza}. }

Another important point is  to understand if one can have a
consistent interacting theory of massless and massive point
particles beyond the cubic level on a flat background. To this end
one can proceed by constructing quartic interaction vertices   and
study their properties
\cite{Sagnotti:2010at}--\cite{Fotopoulos:2010ay},
\cite{Taronna:2011kt}--\cite{Polyakov:2010sk}. The study of quartic
vertices for massless higher spin fields revealed that although the
corresponding quartic Lagrangian is gauge invariant, the symmetry of
the scattering amplitudes requires the inclusion of extra
composite/nonlocal objects into the theory, otherwise the
corresponding S-- matrix is trivial even if the number of fields under
consideration is infinite \cite{Fotopoulos:2010ay} (let us note also
an interesting application of the composite objects in the framework
of the $AdS_4/CFT_3$ correspondence \cite{Jevicki:2012fh}). One  can
therefore  draw the conclusion that gauge invariance itself is not a
sufficient requirement for the consistency of the interacting
theory: rather one should perform some extra tests to investigate if
the interacting  theory is consistent or not.

Apart from the requirement of having nontrivial S-matrix for the
theory of higher spin fields on a flat background, one can employ
the Velo-Zwanziger causality consistency test if the fields under
consideration have non-zero mass. According to \cite{Velo:1969bt} -- \cite{Velo:1972rt}
massive  fields already with spin one interacting with some
nontrivial background can exhibit  noncausal propagation, hence
violating consistency of the theory. Obviously the same kind of
difficulty can appear for massive higher spin fields as well and
checks for different systems have been performed
\cite{Buchbinder:1999ar}--\cite{Buchbinder:2012iz}. In particular in
\cite{Buchbinder:1999ar} it has been shown that the massive fields
with spin 2 propagating on an anti-de Sitter background do not
violate causality. Another famous example of interacting massive
higher spin fields which preserve causality is String Theory. The
Velo-Zwanziger problem in the framework of String Theory has been
considered in detail in \cite{Porrati:2010hm}, where it  was shown
how this problem is avoided.

It is interesting to point out that String Theory is not the only
example of a theory of interacting higher spin fields with non-zero
mass which exhibits causal propagation. The results of
\cite{Buchbinder:2012iz} show that massive higher spin fields
interacting with a background constant electromagnetic field can also
avoid Velo-Zwanziger inconsistency at least at linear order in
electromagnetic field, thus revealing a causal propagation.
 Causality for massive spin $3/2$ coupled to an electromagnetic field
 to all orders in a constant field strength has been proven in \cite{Porrati:2009bs},
and causality for spin $2$ and spin $3/2$ fields on a gravitational background has been 
been discussed in \cite{Buchbinder:2009db}  in the framework of the BRST approach.
 Finally
in the recent paper \cite{Henneaux:2013vca}  it has been argued that
even for a massless theory the gauge invariance itself might not be a
sufficient argument for consistency at the interacting level,
therefore the causality again should be checked separately.

In the present paper we consider the cubic interaction of  massive
higher spin fields with a scalar field. For simplicity we take the
masses of all fields to be the same. The consideration  is actually
performed in two steps. The first step is to check that the
inclusion of the nonlinear cubic interactions into a system which
describes free massive scalar and free massive  higher spin fields
does not change the number of original degrees of freedom. As we
shall see, already this requirement can impose some strong
constraints on the free parameters of interaction and on the mass
parameter. Provided that this requirement is satisfied, the second
step is to perform the causality test for this system. A  completely
rigorous analysis of the causality in the model under consideration
is very complicated. Moreover, it is not very clear  how it can be
done in the case when the number of derivatives in vertices is greater
then the number of
derivatives in a free action.
 In
this paper, we propose a simplified model allowing us to apply the
Velo-Zwanziger procedure for causality analysis. In particular since this
procedure assumes that the number of derivatives in the action is
not higher than two, we shall consider some kind of low-energy
approximation and keep only those terms in the vertex which contain
at most the second derivative acting on the higher spin field.
Besides, the scalar field will be considered as external background.
As a result we get a dynamical higher spin field coupled to an external
scalar field and all derivatives acting on the dynamical field are at
most of the second order. Then we shall see that  causality analysis imposes some additional
requirements on the interaction structure.

We find that it is quite difficult to satisfy both (the preservation
of the correct degrees of freedom and causality) requirements. In
particular the allowed solutions of the first test have been
excluded by the second one and vice versa. The only allowed option
is when the mass parameter $m$ is sufficiently large, which allows
us to ignore certain terms in the original action without imposing
extra conditions on the background. After that we find from the
causality analysis for our simplified model that for certain choices
of coupling constants, or more precisely for certain choices of
parameters entering the cubic interaction vertex, the causality is
preserved, whereas for  the other choices of these parameters the
causality is broken.

To summarize, we proceed as follows. First we consider a relevant cubic Lagrangian
which contains two identical higher spin fields and one scalar. We
consider the scalar to be a background field, whereas the higher
spin field is taken to be dynamical. Then we perform a gauge fixing
procedure to obtain an on-shell cubic  interaction vertex and
 take a low-energy approximation,  i.e. keep only the terms in the Lagrangian which contain a
maximum of two derivatives acting on the dynamical higher spin
fields. After this,  finally we perform the Velo-Zwanziger like
analysis for the system, i.e. compute the characteristic determinant
$D(p)$ and find for which values of the free parameters this
determinant contains the second derivatives acting on the dynamical
field only in the form of the d'Alembertian.

Let us stress that the systems which contain dynamical fields with spin greater then two 
turn out to be the ones where the procedure described above is nontrivial in the following sense.
For the massive fields with spin one and spin two the causality and the correct number of physical degrees of freedom
can be preserved for a constant background scalar field, which in turn means a simple redefinition of the mass parameter in
the theory. For this reason, in the paper we start from the first nontrivial example, i.e. from the 
$3-3-0$ system, and leave a  more detailed discussion of lower spin fields for the Appendices A--B.
In the case of dynamical higher spin fields the 
background scalar is no longer constant, and since it couples to the dynamical higher spin fields
via the derivatives, one has effectively a Lorentz-violating background. This is similar to other examples of
the Velo-Zwanziger problem considered previously in the literature, although in our case 
a nontrivial background scalar field is involved  rather than, say, a nontrivial  electromagnetic background field.

In this paper we will be using the reducible  symmetrical
representations of the Poincar\'{e} group, since an off-shell
formulation for them is  simpler than an off-shell formulation for
irreducible higher spin modes (so--called ``triplet''
\cite{Francia:2002pt}). Therefore below whenever we say ``a
massless triplet with spin $s$'' we actually mean a symmetric tensor
field of rank $s$ along with auxiliary fields with ranks $s-1$ and
$s-2$. The physical polarizations of a triplet contains fields with
spins $s, s-2, ..., 1/0$ with their masses equal to zero. Similarly
a massive triplet (massive reducible representation of the
Poincar\'{e} group) is described by a symmetric field of rank $s$
along with some auxiliary fields. The physical polarizations are
again fields with spins   $s, s-2, ..., 1/0$  with the same value of
mass\footnote{Triplet formulation of higher spin fields can be
further generalized to get completely unconstrained ``quartet''
formulation \cite{Buchbinder:2007}.}.
It would be very  interesting to generalize  our present
analysis for for some other systems such as fields with half integer
spin or fields with mixed symmetries interacting with some nontrivial background \cite{Zinoviev:2010av}
 (see for example
\cite{Buchbinder:2011xw} for recent progress for mixed symmetry
fields). We hope to come back to this issue in future.

The paper is organized as follows.

As a preparation for the massive case, in Section
\ref{3-3-0massless}  we give an explicit example of a massless field
with spin three interacting with a massless scalar in the triplet
formulation. Since we are using an off-shell formulation, the
Lagrangian and equations of motion will contain both physical and
auxiliary fields.
These auxiliary fields, which we denote as $|C\rangle$
and $|D \rangle$, are the feature of the Lagrangian BRST formulations
of the higher spin fields \cite{Koh:1986vg},  \cite{Fotopoulos:2010ay}, \cite{Metsaev:2012uy}--
\cite{Fotopoulos:2007yq} and are absent in the on-shell vertices.
 We present in detail the derivation of the
Lagrangian, of the equations of motion, and of the gauge
transformations for this system, and
show that the number of physical degrees of freedom is preserved after nonlinear deformation
of the free equations.

In Section \ref{3-3-0massive} we carry out an analogous procedure
for the system which contains two identical massive  fields with
spin three and  a massive scalar. After carrying out the gauge
fixing procedure for the  cubic Lagrangian, we find that the
transversality condition can be violated, unless one imposes extra
conditions on the parameters of the theory and on the background. It
means that the gauge invariance in massive case itself does not
guarantee preservation of the correct  number of degrees of freedom.

In Section \ref{3-3-0causality} we consider the aspects of causality.
We formulate a simplified model allowing us to perform the
Velo-Zwanziger like analysis for the $3-3-0$ system and in Section
\ref{s-s-0causality} we generalize these results for the case of
the $s-s-0$ system.

The last section contains our conclusions.

Finally, Appendix \ref{Appendix A} contains detailed expressions for the first--order gauge transformations in
Section \ref{3-3-0massive} and Appendix \ref{Appendix B} contains a discussion of the $1-1-0$ and $2-2-0$ systems.

\section{3-3-0 Vertex: Massless Fields}\label{3-3-0massless}
The goal of this section is to consider some details of a cubic
interaction between two massless spin $3$ triplets with with a
massless scalar. We demonstrate that nonlinear corrections to the
free equations of motion and to the  gauge transformations do not change the
number of physical polarizations. In  other words we check that
in the massless case the requirement of   gauge invariance is sufficient to construct the
cubic vertex which preserves  the correct number of degrees of freedom.

\subsection{Fields and parameters }\label{gen-l}
An off-shell cubic  vertex for
two triplets with spin $3$ and one scalar can be obtained from the vertex (see  \cite{Koh:1986vg},
\cite{Fotopoulos:2010ay} for details)

\begin{equation}\label{V1}
|V \rangle =e^{Y^+_\alpha} e^{Y^+_{gh.}} c_0^{(1)} c_0^{(2)} c_0^{(3)} |0\rangle_{123},
\end{equation}
where
\begin{equation}\label{V2}
Y^+_\alpha= a_1(\alpha^{(1)+} \cdot (p^{(2)} - p^{(3)})  + \alpha^{(2)+} \cdot (p^{(3)} - p^{(1)}) + \alpha^{(3)+} \cdot
(p^{(1)} - p^{(2)})  ),
\end{equation}
\begin{equation}\label{V3}
Y^+_{gh.}=a_1 ( c^{(1)+}(b_0^{(3)} - b_0^{(2)})    +  c^{(2)+}(b_0^{(1)} - b_0^{(3)})  +  c^{(3)+}(b_0^{(2)}- b_0^{(1)})     ).
\end{equation}

Here $a_1$ is an arbitrary  constant and $A \cdot B \equiv A_\mu B^\mu $. The operator $p^{(i)}_\mu$
is a derivative acting on the fields in the $i^{\text{th}}$ Hilbert space. They have the form
   $p^{(i)}_\mu = -i \partial_\mu$ when acting on the right and
 $p^{(i)}_\mu = i \partial_\mu$ when acting on the left.
 The oscillators obey standard (anti-)commutation relations
\begin{equation}
[\alpha_\mu^{(i)}, \alpha_\nu^{(j)+}]= \eta_{\mu \nu} \delta^{ij}, \quad
\{ c_0^{(i)}, b_0^{(j)} \} =  \{ c^{(i)+}, b^{(j)} \} = \{ c^{(i)}, b^{(j)+} \}= \delta^{ij}.
\end{equation}

 Since we are considering an interaction of the type $s-s-0$ the corresponding cubic vertex contains a maximal number of derivatives.
 The vertex (\ref{V2})--(\ref{V3}) has a cyclic symmetry, and so we take the higher spin functional and the parameter of gauge transformations to be of the form.\footnote{In the rest of this subsection no summation over the repeated indices is assumed, unless explicitly specified.}
\begin{eqnarray} \nonumber \label{Phi1}
|\Phi_i \rangle & =& \frac{1}{3!}\phi_{\mu_1 \mu_2 \mu_3} (x) \alpha_{\mu_1}^{(i)+} \alpha_{\mu_2}^{(i)+} \alpha_{\mu_3}^{(i)+}
|0\rangle_i
-
\frac{i}{2!}C_{\mu_1 \mu_2}(x) \alpha_{\mu_1}^{(i)+} \alpha_{\mu_2}^{(i)+}  c^{(i)}_0 b^{(i)+}
|0\rangle_i \\
&&+
D_{\mu_1 }(x) \alpha_{\mu_1}^{(i)+} c^{(i)+} b^{(i)+}
|0\rangle_i
+
\phi(x)|0\rangle_i,
\end{eqnarray}
\be \label{Lambda}
|\Lambda_i \rangle = \frac{i}{2!}\lambda_{\mu_1 \mu_2}(x) \alpha_{\mu_1}^{(i)+} \alpha_{\mu_2}^{(i)+} b^{(i)+}
|0\rangle_i,
\ee
or in a more compact form as
\begin{equation} \label{compact}
|\Phi_i \rangle = |\phi_i \rangle + c^{(i)+} b^{(i)+} | D_i \rangle + c^{(i)}_0 b^{(i)+}| C_i\rangle , \quad
|\Lambda_i \rangle = b^{(i)+}|\lambda_i \rangle.
\end{equation}
The nilpotent BRST charges  for  each Hilbert space are
\begin{equation} \label{brst}
Q^{(i)} =
c_0^{(i)}{l}_0^{(i)} +c^{(i)} l^{(i)+}+c^{(i)+} l^{(i)}  - c^{(i)+}c^{(i)}b^{(i)}_0, \quad i=1,2,3,
\end{equation}
where we used the notation
\begin{equation}
l_0^{(i)} = p^{(i)} \cdot p^{(i)}, \quad  l^{(i)+} = p^{(i)} \cdot \alpha^{(i) +}, \quad  l^{(i)} = p^{(i)} \cdot \alpha^{(i) }.
\end{equation}
Finally the cubic Lagrangian has the form
\be \label{LIBRSTQ}
{L}  =  \sum_{i=1}^3 \int d c_0^{(i)} \langle \Phi_i |\, Q^{(i)} \,|\Phi_i \rangle
  + g( \int dc_0^{(1)} dc_0^{(2)}  dc_0^{(3)} \langle \Phi_1|
\langle \Phi_2|\langle \Phi_3||V_3 \rangle + h.c.).
\ee
The Lagrangian (\ref{LIBRSTQ}) is invariant under the gauge transformations
\begin{equation}\label{transf}
\delta | \Phi_i \rangle = Q^{(i)} |\Lambda_i \rangle
- g
\int
dc_0^{(i+1)} dc_0^{(i+2)}[(  \langle \Phi_{i+1}|\langle \Lambda_{i+2}|
+\langle \Phi_{i+2}|\langle \Lambda_{i+1}|) |V_3 \rangle],
\end{equation}
up to  linear order in the coupling constant $g$
due to the nilpotency of each BRST charge $(Q^{(i)})^2=0$ and the BRST invariance of  vertex
(\ref{V1})
\be
\sum_i^3 Q^{(i)} |V \rangle=0.
\ee

\subsection{Gauge transformations}
In the notation of (\ref{compact}) we have for gauge transformations:
\begin{eqnarray}
&&\delta  |\phi_1 \rangle + c^{(1)+} b^{(1)+} \delta | d_1 \rangle +
 c^{(1)}_0 b^{(1)+} \delta| C_1\rangle
 \\ \nonumber
&&=l^{(1)+} |\lambda _1 \rangle +c_0^{(1)} b^{(1)+} l_0^{(1)}  |\lambda _1 \rangle
+ c^{(1)+} b^{(1)+} l^{(1)} |\lambda _1 \rangle \\ \nonumber
&&+[-ga_1 \langle \phi_2 | \langle \lambda_3|
+ga_1 \langle \phi_3 | \langle \lambda_2|
+ga_1^2 \langle C_3 | \langle \lambda_2|
+ ga_1^2 \langle C_2 | \langle \lambda_3| ] e^{Y^+_\alpha} |0\rangle_{123}.
\end{eqnarray}
Therefore
\begin{equation}\label{trC}
\delta C_{\mu \nu} = \Box \lambda_{\mu \nu},
\end{equation}
\begin{equation}\label{trD}
\delta D_{\mu} = \partial^\nu \lambda_{\mu \nu},
\end{equation}
where $\Box \equiv \partial^\mu \partial_\mu$.

The gauge transformations for $\phi_{\mu_1 \mu_2 \mu_3}$ and $\phi$
are more complicated
\begin{eqnarray} \label{delta3}
&&\left(\frac{1}{3!}\delta \phi_{\mu_1 \mu_2 \mu_3}\alpha^{(1)+}_{\mu_1}\alpha^{(1)+}_{\mu_2}\alpha^{(1)+}_{\mu_3}
+ \delta \phi\right)
|0\rangle_1= \\ \nonumber
 &&\frac{1}{3!}\partial_{(\mu_1}\lambda_{\mu_2 \mu_3)}\alpha^{(1)+}_{\mu_1}\alpha^{(1)+}_{\mu_2}
\alpha^{(1)+}_{\mu_3}|0\rangle_1
+ga_1[- \langle \phi_2 | \langle \lambda_3|
+ \langle \phi_3 | \langle \lambda_2| ] e^{Y^+_\alpha} |0\rangle_{123}.
\end{eqnarray}

 The term proportional to $g$ in
(\ref{delta3}) is a nonabelian deformation of the gauge transformations for
$ \phi_{\mu_1 \mu_2 \mu_3} $
\begin{eqnarray}\label{dphm}
&&- \langle \phi_2 | \langle \lambda_3|   e^{Y^+_\alpha} |0\rangle_{123}
+ \langle \phi_3 | \langle \lambda_2| e^{Y^+_\alpha} |0\rangle_{123}= \\ \nonumber
&&+\prescript{}{2}\langle 0|  \prescript{}{3}\langle 0| \phi  \frac{i}{2}(\lambda_{\nu_1 \nu_2} \alpha_{\nu_1}^{(3)}
  \alpha_{\nu_2}^{(3)})
\frac{1}{2} {(a_1\alpha^{(3)+} \!\cdot (p^{(1)}\! -p^{(2)}))}^2 \nonumber \\
&& \hspace{45mm}\times\frac{1}{3!} {(a_1\alpha^{(1)+} \!\cdot (p^{(2)}\! -p^{(3)}))}^3
 |0\rangle_{123} \nonumber
\\ \nonumber
&&-\prescript{}{3}\langle 0|  \prescript{}{2}\langle 0| \phi  \frac{i}{2}(\lambda_{\nu_1 \nu_2} \alpha_{\nu_1}^{(2)}
 \alpha_{\nu_2}^{(2)})
\frac{1}{2} {(a_1\alpha^{(2)+} \!\cdot (p^{(3)}\! -p^{(1)}))}^2 \nonumber \\
&& \hspace{45mm}\times\frac{1}{3!} {(a_1\alpha^{(1)+} \!\cdot
 (p^{(2)}\! -p^{(3)}))}^3  |0\rangle_{123}. \nonumber
\end{eqnarray}

The operator $p_\mu^{(1)}$  in the equation  (\ref{dphm})  should be replaced with $-p_\mu^{(2)} - p_\mu^{(3)}$ due to
 the relation
\begin{equation}
p_\mu^{(1)} + p_\mu^{(2)} + p_\mu^{(3)}=0,
\end{equation}
which reflects the fact that one can discard the total derivative in the Lagrangian.
This is justified in the equations of motion and gauge transformation
rules since they are Lagrangian equations and represent invariance of a Lagrangian.

Finally one obtains for the   tensor field
\begin{eqnarray}\label{dp3}
&&\frac{1}{3!}\delta \phi_{\mu_1 \mu_2 \mu_3}= \frac{1}{3!}\partial_{(\mu_1}\lambda_{\mu_2 \mu_3)} - \\ \nonumber
&& \frac{ga_1^6}{2}\left[4 (\partial_{\mu_1 \mu_2 \mu_3 \nu_1 \nu_2} \phi) \lambda_{\nu_1 \nu_2}
+ 4 (\partial_{\mu_1 \mu_2 \mu_3 \nu_1 } \phi)(\partial_{\nu_2}   \lambda_{\nu_1 \nu_2})
+(\partial_{\mu_1 \mu_2 \mu_3 } \phi)(\partial_{\nu_1 \nu_2}   \lambda_{\nu_1 \nu_2})\right. \\ \nonumber
&& -12
(\partial_{\mu_1 \mu_2 \nu_1 \nu_2} \phi)(\partial_{\mu_3}   \lambda_{\nu_1 \nu_2})
-12
(\partial_{\mu_1 \mu_2 \nu_1 } \phi)(\partial_{\nu_2 \mu_3}   \lambda_{\nu_1 \nu_2})
-3
(\partial_{\mu_1 \mu_2  } \phi)(\partial_{\nu_1 \nu_2 \mu_3}   \lambda_{\nu_1 \nu_2}) \\ \nonumber
&&
+12(\partial_{\mu_1 \nu_1 \nu_2 } \phi)(\partial_{\mu_2 \mu_3}   \lambda_{\nu_1 \nu_2})
+12(\partial_{\mu_1 \nu_1  } \phi)(\partial_{\mu_2 \mu_3 \nu_2}   \lambda_{\nu_1 \nu_2})
+ 3(\partial_{\mu_1  } \phi)(\partial_{\mu_2 \mu_3 \nu_1 \nu_2}   \lambda_{\nu_1 \nu_2}) \\ \nonumber
&&
\left. -4
(\partial_{\nu_1 \nu_2  } \phi)(\partial_{\mu_1 \mu_2 \mu_3}   \lambda_{\nu_1 \nu_2})
-4
(\partial_{\nu_1   } \phi)(\partial_{\mu_1 \mu_2 \mu_3 \nu_2}   \lambda_{\nu_1 \nu_2})
-
 \phi(\partial_{\mu_1 \mu_2 \mu_3  \nu_1 \nu_2}   \lambda_{\nu_1 \nu_2})\right],
\end{eqnarray}
where the nonlinear terms on the right hand--side of the equation (\ref{dp3}),  as well
as in all analogous equations below,
are assumed to be  symmetrized with weight 1 with respect to the free indices.
Similarly for the scalar one has
\begin{eqnarray}
&&\delta \phi =  -\frac{ga_1^6}{3!}\left[32 (\partial_{\rho_1 \rho_2} \phi_{\nu_1 \nu_2 \nu_3})(\partial_{\nu_1 \nu_2 \nu_3} \lambda_{\rho_1 \rho_2})
+
32 (\partial_{\rho_1 } \phi_{\nu_1 \nu_2 \nu_3})(\partial_{\nu_1 \nu_2 \nu_3 \rho_2} \lambda_{\rho_1 \rho_2})\right. \nonumber \\
&&
+
8 \,\phi_{\nu_1 \nu_2 \nu_3}(\partial_{\nu_1 \nu_2 \nu_3  \rho_1 \rho_2} \lambda_{\rho_1 \rho_2})
+48
 (\partial_{\nu_1 \rho_1 \rho_2} \phi_{\nu_1 \nu_2 \nu_3})(\partial_{ \nu_2 \nu_3} \lambda_{\rho_1 \rho_2})
 \nonumber \\
&&
+ 48
(\partial_{\nu_1 \rho_1 } \phi_{\nu_1 \nu_2 \nu_3})(\partial_{ \nu_2 \nu_3 \rho_2} \lambda_{\rho_1 \rho_2})
+12
(\partial_{\nu_1  } \phi_{\nu_1 \nu_2 \nu_3})(\partial_{ \nu_2 \nu_3  \rho_1 \rho_2} \lambda_{\rho_1 \rho_2})
\nonumber \\
&&+ 24
 (\partial_{\nu_1 \nu_2 \rho_1 \rho_2} \phi_{\nu_1 \nu_2 \nu_3})(\partial_{ \nu_3} \lambda_{\rho_1 \rho_2})
+24
 (\partial_{\nu_1 \nu_2 \rho_1 } \phi_{\nu_1 \nu_2 \nu_3})(\partial_{ \nu_3 \rho_2} \lambda_{\rho_1 \rho_2})
\nonumber \\
&&
+6
 (\partial_{\nu_1 \nu_2 } \phi_{\nu_1 \nu_2 \nu_3})(\partial_{ \nu_3 \rho_1 \rho_2} \lambda_{\rho_1 \rho_2})
+4
 (\partial_{\nu_1 \nu_2 \nu_3 \rho_1 \rho_2} \phi_{\nu_1 \nu_2 \nu_3}) \lambda_{\rho_1 \rho_2} \nonumber \\
 &&
\left. + 4
 (\partial_{\nu_1 \nu_2  \nu_3 \rho_1 } \phi_{\nu_1 \nu_2 \nu_3})(\partial_{ \rho_2} \lambda_{\rho_1 \rho_2})
+
 (\partial_{\nu_1 \nu_2  \nu_3  } \phi_{\nu_1 \nu_2 \nu_3})(\partial_{\rho_1 \rho_2} \lambda_{\rho_1 \rho_2})\right]
\nonumber \\
&& +\frac{ga_1^6}{2} \left[16 ( \partial_{\mu_1 \mu_2} \lambda_{\nu_1 \nu_2})(\partial_{\nu_1 \nu_2}C_{\mu_1 \mu_2})
+ 16
(\partial_{\mu_1 \mu_2 \nu_1} \lambda_{\nu_1 \nu_2})(\partial_{ \nu_2}C_{\mu_1 \mu_2})\right. \nonumber \\
&& +4
(\partial_{\mu_1 \mu_2 \nu_1 \nu_2} \lambda_{\nu_1 \nu_2})C_{\mu_1 \mu_2}
 + 16
(\partial_{\mu_1 } \lambda_{\nu_1 \nu_2})(\partial_{\mu_2 \nu_1 \nu_2 }C_{\mu_1 \mu_2}) \nonumber \\
&&
+
16(
\partial_{\mu_1 \nu_1 } \lambda_{\nu_1 \nu_2})(\partial_{\mu_2  \nu_2 }C_{\mu_1 \mu_2})
+4(
\partial_{\mu_1 \nu_1 \nu_2 } \lambda_{\nu_1 \nu_2})(\partial_{\mu_2   }C_{\mu_1 \mu_2})
\nonumber \\
&&
\left. +4  \lambda_{\nu_1 \nu_2} (\partial_{\mu_1 \mu_2 \nu_1 \nu_2}C_{\mu_1 \mu_2})
+4 (\partial_{\nu_2}\lambda_{\nu_1 \nu_2}) (\partial_{\mu_1 \mu_2 \nu_1 }C_{\mu_1 \mu_2})
 +(\partial_{\nu_2 \nu_2}\lambda_{\nu_1 \nu_2}) (\partial_{\mu_1 \mu_2  }C_{\mu_1 \mu_2})\right]. \nonumber \\
\end{eqnarray}

\subsection{Equations of Motion}
From the Lagrangian (\ref{LIBRSTQ})
written in terms of component fields (\ref{compact}), i.e. after integrating out the ghost variables
\bea
&& L=
\sum_{i=1,2,3}\left(
\langle \phi_i|l_0^{(i)}|\phi_i \rangle - \langle D_i|l_0^{(i)}|D_i \rangle +
\langle C_i||C_i \rangle - \langle \phi^i | l^{(i)+}  |C^i \rangle
+  \langle D_i|l^{(i)}|C_i \rangle \right. \nonumber \\
&& \hspace{20mm}\left.
- \langle C_i|l^{(i)}|\phi_i \rangle
+  \langle C_i|l^{(i)+}|D_i \rangle\right) \\
&& -g\left[\left(
\langle \phi_3| \langle \phi_2| \langle \phi_1|
+
a_1^2\langle C_3 | \langle C_2| \langle \phi_1 |
+
 a_1^2\langle C_2 | \langle C_1| \langle \phi_3 |
+
a_1^2\langle C_3 | \langle C_1 |\langle \phi_2 |\right) e^{Y^+_\alpha}   |0\rangle_{123}\right. \nonumber \\
&&\left. \hspace{10mm} +
h.c\right], \nonumber
\end{eqnarray}
one can readily derive the corresponding equations of motion.

The equation of motion with respect to
 $\langle \phi_1|$ :
\begin{equation} \label{phi}
l_0^{(1)}|\phi_1 \rangle -  l^{(1)+}  |C_1 \rangle - g \langle \phi_3| \langle \phi_2| e^{Y^+_\alpha}   |0\rangle_{123}
-g a_1^2\langle C_3 | \langle C_2| e^{Y^+_\alpha}   |0\rangle_{123} =0.
\end{equation}

Let us note that this expression actually  contains two equations: one is with respect to
$\phi_{\mu_1 \mu_2 \mu_3}$ and the other is with respect to $\phi$.

The equation of motion with respect to  $\langle C_1 |$:
\begin{equation}\label{C}
|C_1 \rangle - l^{(1)}|\phi_1\rangle + l^{(1)+}|D_1\rangle
 -g a_1^2
( \langle C_2 |  \langle \phi_3 |
+
\langle C_3 |  \langle \phi_2 |) e^{Y^+_\alpha}   |0\rangle_{123} =0,
\end{equation}
and finally the equation of motion with respect to $\langle D_1|$:

\begin{equation}\label{D}
l_0^{(1)}|D_1 \rangle-   l^{(1)}  |C_1 \rangle=0.
\end{equation}

Let us first consider the equation with respect to $\phi_{\mu_1 \mu_2 \mu_3}$.
It contains two parts:
\begin{equation}\label{ephi3-1}
l_0^{(1)}|\phi_1 \rangle -  l^{(1)+}  |C_1 \rangle = \left[-\Box \phi_{\nu_1 \nu_2 \nu_3}
+ \partial_{(\nu_1}C_{\nu_2 \nu_3)}\right]\frac{1}{3!}\alpha^{(1)+}_{\nu_1}\alpha^{(1)+}_{\nu_2}\alpha^{(1)+}_{\nu_3}|0\rangle_1,
\end{equation}
and
\bea\label{ephi3-2}
 &&- g \langle \phi_3| \langle \phi_2| e^{Y^+_\alpha}   |0\rangle_{123}= \\ \nonumber
&&+\frac{ga_1^6}{3!}\left[16(\partial_{\nu_1 \nu_2 \nu_3} \phi_{\mu_1 \mu_2 \mu_3}) (\partial_{\mu_1 \mu_2 \mu_3} \phi)
-48 (\partial_{\nu_1 \nu_2 } \phi_{\mu_1 \mu_2 \mu_3}) (\partial_{\mu_1 \mu_2 \mu_3 \nu_3} \phi)\right. \\ \nonumber
&&+ 48
(\partial_{\nu_1  } \phi_{\mu_1 \mu_2 \mu_3}) (\partial_{\mu_1 \mu_2 \mu_3  \nu_2 \nu_3} \phi)
-16
\phi_{\mu_1 \mu_2 \mu_3} (\partial_{\mu_1 \mu_2 \mu_3  \nu_1 \nu_2 \nu_3} \phi) \\ \nonumber
&&+24
(\partial_{\mu_1 \nu_1 \nu_2 \nu_3} \phi_{\mu_1 \mu_2 \mu_3}) (\partial_{ \mu_2 \mu_3} \phi)
-72
(\partial_{\mu_1 \nu_1 \nu_2 } \phi_{\mu_1 \mu_2 \mu_3 }) (\partial_{ \mu_2 \mu_3 \nu_3} \phi) \\ \nonumber
&&+72
(\partial_{\mu_1 \nu_1  } \phi_{\mu_1 \mu_2 \mu_3 }) (\partial_{\mu_2 \mu_3 \nu_2 \nu_3} \phi)
 -24
(\partial_{\mu_1 } \phi_{\mu_1 \mu_2 \mu_3 }) (\partial_{\mu_2 \mu_3 \nu_1 \nu_2 \nu_3} \phi) \\ \nonumber
&&+12(\partial_{\mu_1 \mu_2 \nu_1 \nu_2 \nu_3} \phi_{\mu_1 \mu_2 \mu_3})(\partial_{\mu_3} \phi)
-36
(\partial_{\mu_1 \mu_2 \nu_1 \nu_2 } \phi_{\mu_1 \mu_2 \mu_3})(\partial_{\mu_3 \nu_3} \phi) \\ \nonumber
&& +36
(\partial_{\mu_1 \mu_2 \nu_1  } \phi_{\mu_1 \mu_2 \mu_3})(\partial_{\mu_3 \nu_2 \nu_3} \phi)
-12
(\partial_{\mu_1 \mu_2   } \phi_{\mu_1 \mu_2 \mu_3})(\partial_{\mu_3 \nu_1 \nu_2 \nu_3} \phi) \\ \nonumber
&& +2(\partial_{\mu_1 \mu_2 \mu_3 \nu_1 \nu_2 \nu_3} \phi_{\mu_1 \mu_2 \mu_3}) \phi
-6
 (\partial_{\mu_1 \mu_2 \mu_3 \nu_1 \nu_2 } \phi_{\mu_1 \mu_2 \mu_3})( \partial_{\nu_3} \phi) \\ \nonumber
&&\left. +6
 (\partial_{\mu_1 \mu_2 \mu_3 \nu_1  } \phi_{\mu_1 \mu_2 \mu_3})( \partial_{\nu_2\nu_3} \phi)
-
 2(\partial_{\mu_1 \mu_2 \mu_3   } \phi_{\mu_1 \mu_2 \mu_3})( \partial_{\nu_1 \nu_2\nu_3} \phi)\right]\alpha^{(1)+}_{\nu_1}\alpha^{(1)+}_{\nu_2}\alpha^{(1)+}_{\nu_3}|0\rangle_1.
\eea
Now let us turn to the equations with respect to
$\phi$.
It consists from the following parts:
\begin{equation}\label{ephi0-1}
l_0^{(1)}|\phi_1 \rangle  = -\Box \phi|0\rangle_1,
\end{equation}
as well as
\bea \label{ephi0-2}
 && g \langle \phi_3| \langle \phi_2| e^{Y^+_\alpha}   |0\rangle_{123}= \\ \nonumber
&&+\frac{ga_1^6}{3! 3!}\left[16 \phi_{\mu_1 \mu_2 \mu_3}(\partial_{\mu_1 \mu_2 \mu_3 \nu_1 \nu_2 \nu_3} \phi_{\nu_1 \nu_2 \nu_3})
+ 24(\partial_{\mu_1} \phi_{\mu_1 \mu_2 \mu_3})(\partial_{ \mu_2 \mu_3 \nu_1 \nu_2 \nu_3} \phi_{\nu_1 \nu_2 \nu_3})\right. \\ \nonumber
&&+12
(\partial_{\mu_1 \mu_2 \mu_3 \nu_1} \phi_{\mu_1 \mu_2 \mu_3})(\partial_{  \nu_2 \nu_3} \phi_{\nu_1 \nu_2 \nu_3})
+192
(\partial_{\mu_3 \nu_1 \nu_2 \nu_3} \phi_{\mu_1 \mu_2 \mu_3})(\partial_{  \mu_1 \mu_2} \phi_{\nu_1 \nu_2 \nu_3}) \\
\nonumber
&&+ 144
(\partial_{\mu_1 \nu_1 } \phi_{\mu_1 \mu_2 \mu_3})(\partial_{  \mu_2 \mu_3 \nu_2 \nu_3} \phi_{\nu_1 \nu_2 \nu_3})
+ 96
(\partial_{\mu_1  \mu_2 \nu_1 \nu_2 \nu_3 } \phi_{\mu_1 \mu_2 \mu_3})(\partial_{\mu_3 } \phi_{\nu_1 \nu_2 \nu_3}) \\
\nonumber
&&+ 36
(\partial_{\mu_1  \mu_2 \nu_1  } \phi_{\mu_1 \mu_2 \mu_3})(\partial_{\mu_3 \nu_2 \nu_3 } \phi_{\nu_1 \nu_2 \nu_3})
+ 144
(\partial_{\mu_3  \nu_2 \nu_3  } \phi_{\mu_1 \mu_2 \mu_3})(\partial_{\mu_1 \mu_2 \nu_1 } \phi_{\nu_1 \nu_2 \nu_3})
 \\
\nonumber
&&\left. + (\partial_{\mu_1  \mu_2 \mu_3  } \phi_{\mu_1 \mu_2 \mu_3})(\partial_{\nu_1 \nu_2 \nu_3 } \phi_{\nu_1 \nu_2 \nu_3})
+ 64
(\partial_{\nu_1  \nu_2 \nu_3  } \phi_{\mu_1 \mu_2 \mu_3})(\partial_{\mu_1 \mu_2 \mu_3 } \phi_{\nu_1 \nu_2 \nu_3})\right]|0\rangle_1 ,
\eea
and
\bea
&&-ga_1^2 \langle C_3 | \langle C_2| e^{Y^+_\alpha}   |0\rangle_{123} =  \frac{ga_1^6}{4}\left[8 (\partial_{\mu_1 \mu_2 \nu_1 \nu_2}C_{\mu_1 \mu_2})C_{\nu_1 \nu_2} \right. \nonumber \\
&&
+
32  (\partial_{\nu_1}C_{\mu_1 \mu_2})(\partial_{\nu_2 \mu_1 \mu_2}C_{\nu_1 \nu_2})+ 8 (\partial_{\mu_1}C_{\mu_1 \mu_2})(\partial_{\mu_2 \nu_1 \nu_2}C_{\nu_1 \nu_2})
 \nonumber \\
&&
+16(\partial_{\nu_1 \nu_2}C_{\mu_1 \mu_2})(\partial_{\mu_1 \mu_2 }C_{\nu_1 \nu_2})
+16(\partial_{\mu_1 \nu_1}C_{\mu_1 \mu_2})(\partial_{\mu_2 \nu_2 }C_{\nu_1 \nu_2}) \nonumber \\
&& \left. +
(\partial_{\mu_1 \mu_2}C_{\mu_1 \mu_2})(\partial_{\nu_1 \nu_2 }C_{\nu_1 \nu_2})\right]|0\rangle_1 .
\eea

Let us turn to the equation of motion with respect to
$\langle C_1|$.
It contains parts
\begin{eqnarray}\label{eC1}
&&|C_1 \rangle - l^{(1)}|\phi_1\rangle + l^{(1)+}|D_1\rangle= \\ \nonumber
&&\left[-i
C_{\nu_1 \nu_2} +i\partial^{\mu_1}\phi_{\mu_1 \nu_1 \nu_2} -i \partial_{(\nu_1}D_{\nu_2)}\right]\frac{1}{2!}\alpha^{(1)+}_{\nu_1}\alpha^{(1)+}_{\nu_2}|0\rangle_1,
\end{eqnarray}

\begin{eqnarray}\label{eC2}
&&-g a_1^2
( \langle C_2 |  \langle \phi_3 |
+
\langle C_3 |  |\langle \phi_2 |) e^{Y^+_\alpha}   |0\rangle_{123} = \\ \nonumber
&& \frac{iga_1^6}{2}\left[ 4(\partial_{ \nu_1 \nu_2}C_{\mu_1 \mu_2}) (\partial_{\mu_1 \mu_2}\phi) -
8(\partial_{ \nu_1 }C_{\mu_1 \mu_2}) (\partial_{\nu_2 \mu_1 \mu_2} \phi)
+ 4
C_{\mu_1 \mu_2} (\partial_{\mu_1 \mu_2 \nu_1 \nu_2} \phi)\right. \\ \nonumber
&& +4
(\partial_{ \mu_1 \nu_1 \nu_2 }C_{\mu_1 \mu_2}) (\partial_{\mu_2 } \phi) -
8 (\partial_{ \mu_1 \nu_1  }C_{\mu_1 \mu_2}) (\partial_{\mu_2 \nu_2 } \phi)
+ 4  (\partial_{ \mu_1   }C_{\mu_1 \mu_2}) (\partial_{\mu_2 \nu_1 \nu_2 } \phi) \\ \nonumber
&& \left. +(\partial_{ \mu_1 \mu_2 \nu_1 \nu_2  }C_{\mu_1 \mu_2})  \phi- 2
 (\partial_{ \mu_1 \mu_2 \nu_1   }C_{\mu_1 \mu_2}) (\partial_{ \nu_2 } \phi)
+(\partial_{\mu_1 \mu_2} C_{\mu_1 \mu_2}) (\partial_{ \nu_1 \nu_2 } \phi)\right]\alpha^{1+}_{\nu_1}\alpha^{2+}_{\nu_2}|0\rangle_1.
\end{eqnarray}
Finally equations of motion
with respect to $\langle D_1|$ are
\begin{equation}\label{eD}
\Box D_\mu = \partial^\nu C_{\mu \nu},
\end{equation}
so the equation of motion with respect to the field
$\langle D_1|$ is not modified by nonlinear terms.

\subsection{Gauge fixing}\label{gfpm}

Let us first recall how the light--cone gauge fixing procedure can be performed for a triplet in
 the absence of interactions,  i.e. when $g=0$.
To this end one can use the parameter $\lambda_{\mu \nu}$ in (\ref{dp3}) to eliminate
$\phi_{+++}, \phi_{+ij}, \phi_{++i}, \phi_{+i-}, \phi_{++-}, \phi_{+--}, ( i,j=1,...,{\cal D}-2$) components from the
field $\phi_{\mu_1 \mu_2 \mu_3}$.
After we have used the entire gauge freedom we can turn to the equations of motion.
The components  of the equation of motion (\ref{phi}) which contain at least one index ``+''
imply that the field $C_{\mu_1 \mu_2}$ is zero.
In the same way the components of the equation (\ref{C}) which contain at least one
index ``+''  implies that the field $D_\mu$  is zero, thus turning the equation ($\ref{D}$)
into an identity.

The rest of the components of the equation (\ref{C}) implies the transversality of the fields $\phi_{\mu_1 \mu_2 \mu_3}$
\begin{equation}\label{tr-m}
\partial^{\mu_1}\phi_{\mu_1 \mu_2 \mu_3}=0,
\end{equation}
the latter condition eliminating the components of the field $\phi_{\mu_1 \mu_2 \mu_3}$ which contains the index
``-". Finally, the equation (\ref{phi})  gives the mass-shell condition for the longitudinal components
\begin{equation}
\Box \phi_{i_1 i_2 i_3}=0.
\end{equation}

Let us note that this gauge fixing procedure is valid for an arbitrary number of space--time dimensions and for an arbitrary value
of the spin of the triplet \cite{Francia:2002pt}.

The modification of this procedure to the case of non-zero coupling constant is straightforward.
One can check that after imposing light cone gauge on the field $\phi_{\mu_1 \mu_2 \mu_3 }$
the components of the nonlinear equation (\ref{phi}) which contain at least one index ``+"
still imply that the field $C_{\mu_1 \mu_2}$ vanishes.
Therefore one ends up with the nonlinear equations
\bea
&& -\frac{1}{3!}\Box \phi_{i_1 i_2 i_3 }
+\frac{ga_1^6}{3!}\left[16(\partial_{i_1 i_2 i_3} \phi_{j_1 j_2 j_3}) (\partial_{j_1 j_2 j_3} \phi)
-48 (\partial_{i_1 i_2 } \phi_{j_1 j_2 j_3}) (\partial_{ j_1 j_2 j_3 i_3} \phi)\right. \nonumber \\
&&\left. + 48
(\partial_{i_1  } \phi_{ j_1 j_2 j_3}) (\partial_{ j_1 j_2 j_3  i_2 i_3} \phi)
-16
\phi_{ j_1 j_2 j_3} (\partial_{ j_1 j_2 j_3  i_1 i_2 i_3} \phi)\right] =0,
\eea
and
\be
-\Box \phi + \frac{64ga_1^6}{3! 3!}
 (\partial_{i_1  i_2 i_3  } \phi_{ j_1 j_2 j_3})(\partial_{ j_1 j_2 j_3} \phi_{i_1 i_2 i_3})=0.
\ee

The results of this section can be summarized as follows. For
massless higher spin fields one first constructs a free Lagrangian
(the first term in (\ref{LIBRSTQ})) using a nilpotent BRST charge
(\ref{brst}). Then using the linear gauge transformations  (the
first term in (\ref{transf})) and free equations of motion one can
gauge away all auxiliary fields present in the free Lagrangian. The
next step is to construct a cubic Lagrangian and to make a nonlinear
deformation of the initial linear gauge transformations in such a way
that that the number of gauge parameters are preserved. Here we have
obtained all relevant equations explicitly for the system $3-3-0$
and finally demonstrated that after the total gauge fixing the
number of propagating degrees of freedom, which now obey nonlinear
equations of motion,  is preserved.

Although these results might have been anticipated for the case of
massless fields we believe that this explicit consideration is
useful especially as a  preparation for the case of massive fields,
where the situation is completely different.

\setcounter{equation}0
\section{3-3-0 Vertex: Massive Fields}\label{3-3-0massive}
In this Section we consider in detail the structure of the cubic
interaction of and equations of motion for two massive spin $3$
triplets coupled to a massive scalar. As in the case of massless
higher spin fields we again have some free parameters in the BRST
invariant vertex. The goal is to consider the gauge fixing procedure
at the nonlinear (cubic) level and study whether the gauge invariance
imposes such strong restrictions on the parameters as in the massless
case.

\subsection{Fields and Parameters}

The general line for the construction of the cubic Lagrangians for massive fields is the same
as that given in Section \ref{gen-l}, with a few differences to be discussed below.

The cubic Lagrangian \eqref{LIBRSTQ} and gauge transformations \eqref{transf} have the same form as for the massless fields.
 However, the nilpotent BRST charge $Q^{(i)}$, which can be obtained from the dimensional reduction
from a $({\cal D}+1)$-dimensional massless theory, i.e. from the charge
 \eqref{brst}, is given by
 \cite{Hussain:1988uk}--\cite{Pashnev:1997rm}\footnote{
 Free higher spin bosonic Lagrangian theory can also be formulated
 on the base of BRST construction without dimensional reduction both
 in flat and in AdS spaces \cite{Buchbinder:2005}.}
\begin{equation}\label{massbrst}
Q^{(i)}=c_0^{(i)}(l_0^{(i)}+m_i^2)+c^{(i)+}(l^{(i)}+m_i\alpha^{i}_D)+c^{(i)}(l^{(i)+}+m_i\alpha^{(i)+}_D)-c^{(i)+}
c^{(i)}b_0^{(i)}, \quad i=1,2,3,
\end{equation}
where $m_i$ are the masses of the fields in the $i^{th}$ Hilbert space. An extra oscillator
\begin{equation}
[\alpha_D^{(i)}, \alpha_D^{(j)+}]= \delta^{ij},
\end{equation}
corresponds to the reduced dimension.
 Let us note that
the mass parameter is not necessarily a constant, rather it can be a
function of the spin, thus describing a Regge trajectory
\cite{Pashnev:1997rm}.

An off-shell cubic vertex for the massive higher spin fields with different masses was given in
\cite{Metsaev:2012uy} in terms of the BRST closed forms
\begin{equation}\label{f-1}
L^{(i)}=a_1\left(\alpha^{(i)} \cdot (p^{(i+1)}-p^{(i+2)}) -c^{(i)}(b_0^{(i+2)}-b_0^{(i+1)}) -\frac{m_{i+1}^2-m_{i+2}^2}{m_i}\alpha^{(i)}_D\right),
\end{equation}
and
\begin{eqnarray}\label{f-2}
Q^{(i,i+1)}&=& a_2\left(\alpha^{(i)} \cdot \alpha^{(i+1)\,} +\frac{\alpha^{(i)}_D}{2a_1m_i} L^{(i+1)}-\frac{\alpha^{(i+1)}_D}{2a_1m_{i+1}} L^{(i)} \right. \\
&& \left. \hspace{8mm}-\frac{m_i^2+m_{i+1}^2-m_{i+2}^2}{2m_im_{i+1}}\alpha^{(i)}_D\alpha^{(i+1)}_D -\frac{1}{2}b^{(i)}c^{(i+1)\,}-\frac{1}{2}b^{(i+1)\,}c^{(i)}\right), \nonumber
\end{eqnarray}
where $a_1$ and $a_2$ are arbitrary real constants.
Similarly to the cubic vertex for massless higher spin fields, since the expressions (\ref{f-1})
and (\ref{f-2}) are separately BRST invariant, any function of these expressions is BRST invariant as well.

The higher spin functionals for the massive triplet can be deduced
 from the dimensional reduction of the massless triplets. In particular,  for the spin $3$
triplet we have
\begin{align} \nonumber \label{massPhi1}
|\Phi_{1,2} \rangle & = \frac{1}{3!}\phi_{\mu_1 \mu_2 \mu_3} (x) \alpha_{\mu_1}^{(1,2)+} \alpha_{\mu_2}^{(1,2)+} \alpha_{\mu_3}^{(1,2)+}
|0\rangle_{1,2}
+\frac{i}{2!}h_{\mu_1\mu_2}(x)\alpha_{\mu_1}^{(1,2)+} \alpha_{\mu_2}^{(1,2)+} \alpha_{D}^{(1,2)+}|0\rangle_{1,2} \\ \nonumber
&+b_{\mu_1}(x)\alpha_{\mu_1}^{(1,2)+} \alpha_{D}^{(1,2)+} \alpha_{D}^{(1,2)+}|0\rangle_{1,2}
+i\varphi(x)\alpha_{D}^{(1,2)+} \alpha_{D}^{(1,2)+} \alpha_{D}^{(1,2)+}|0\rangle_{1,2} \\ \nonumber
&-
\frac{i}{2!}C_{\mu_1 \mu_2}(x) \alpha_{\mu_1}^{(1,2)+} \alpha_{\mu_2}^{(1,2)+}  c^{(1,2)}_0 b^{(1,2)+}
|0\rangle_{1,2}  \\ \nonumber
&
-
C_{\mu_1}(x) \alpha_{\mu_1}^{(1,2)+} \alpha_{D}^{(1,2)+}  c^{(1,2)}_0 b^{(1,2)+}
|0\rangle_{1,2}
-
iC(x) \alpha_{D}^{(1,2)+} \alpha_{D}^{(1,2)+}  c^{(1,2)}_0 b^{(1,2)+}
|0\rangle_{1,2} \\
&+
D_{\mu_1 }(x) \alpha_{\mu_1}^{(1,2)+} c^{(1,2)+} b^{(1,2)+}
|0\rangle_{1,2}
+
iD(x) \alpha_{D}^{(1,2)+} c^{(1,2)+} b^{(1,2)+}
|0\rangle_{1,2},
\end{align}
and
\begin{equation} \label{massPhi3}
|\Phi_3\rangle = \phi(x)|0\rangle_3.
\end{equation}

Similarly, parameters of the gauge transformations take the form
\begin{eqnarray}\label{massLambda1}
|\Lambda_{1,2}\rangle &=& \frac{i}{2!}\lambda_{\mu_1\mu_2}(x)\alpha_{\mu_1}^{(1,2)+}\alpha_{\mu_2}^{(1,2)+}b^{(1,2)+}
|0\rangle_{1,2}
+
\lambda_{\mu_1}(x)\alpha_{\mu_1}^{(1,2)+}\alpha_{D}^{(1,2)+}b^{(1,2)+}|0\rangle_{1,2} \nonumber \\
&&+
i\lambda(x)\alpha_{D}^{(1,2)+}\alpha_{D}^{(1,2)+}b^{(1,2)+}|0\rangle_{1,2},
\end{eqnarray}
and
\begin{equation}
|\Lambda_3\rangle=0.
\end{equation}

As we shall see below when discussing the gauge fixing procedure the fields $ h_{\mu_1\mu_2}, b_{\mu_1}, \varphi $,
are St\"{u}ckelberg fields, which are required for the gauge invariant description of massive fields,
whereas the fields $C_{\mu_1\mu_2}, C_{\mu_1}, C,  D_{\mu_1}$ and $D$ are auxiliary fields, similar
to the ones  that are present  in the description of the massless triplet.

In the case of the $3-3-0$ vertex the full expression is given by
\begin{eqnarray}\label{masscubic}
|V\rangle &=&\left[\frac{1}{3!3!}(L^{(1)+})^3(L^{(2)+})^3+\frac{1}{2!2!}(L^{(1)+})^2(L^{(2)+})^2Q^{(12)+}  \right. \nonumber \\
&&\hspace{10mm}\left.
+
\frac{1}{2!}L^{(1)+}L^{(2)+}(Q^{(12)+})^2+\frac{1}{3!}(Q^{(12)+})^3 \right] c_0^{(1)}c_0^{(2)}c_0^{(3)}|0\rangle_{123},
\end{eqnarray}
where the operators $L^{(1,2)+}$ and $Q^{(12)+}$ are hermitian conjugate to the operators
(\ref{f-1}) and (\ref{f-2}).
We will  refer to various parts of the vertex (\ref{masscubic}) as
\be\label{V6V5}
V_6:=\frac{1}{3!3!}(L^{(1)+})^3(L^{(2)+})^3, \hspace{3mm} V_5:=\frac{1}{2!2!}(L^{(1)+})^2(L^{(2)+})^2Q^{(12)+},
\ee
\be\label{V4V3}
 V_4:=\frac{1}{2!}L^{(1)+}L^{(2)+}(Q^{(12)+})^2, \hspace{3mm} V_3:=\frac{1}{3!}(Q^{(12)+})^3,
\ee
where the subscripts denote the maximal number of derivatives which will appear in the corresponding Lagrangian.
It is easiest when computing the expressions for the Lagrangian and gauge transformations to consider the contributions from each of these vertices separately.

 Let us note that since we are considering the identical triplets in the first and in the second Hilbert spaces,
one has $m_1=m_2$. Further considerable simplification occurs when the mass of the scalar field is
equal to the mass of the triplet. Therefore below we will consider the situation when  $m_1=m_2=m_3:=m$.

Another important point is that keeping in mind the subsequent
application of the Velo-Zwanziger procedure we shall consider only
the $V_6$ and $V_5$ parts (\ref{V6V5}) of the vertex. The $V_4$ and
$V_3$ parts (\ref{V4V3}) of the vertex give rise to terms with fewer
then two derivatives on the dynamical field and are therefore
irrelevant for the causality analysis.

\subsection{Gauge Transformations}
The gauge transformations to zeroth--order in $g$ read
\be
\delta_0|\Phi_i\rangle=Q^{(i)}|\Lambda_i\rangle,
\ee
where $Q^i$ are as in \eqref{massbrst}, which gives rise to \cite{Pashnev:1997rm}
\begin{eqnarray}\label{brstmtr}
\delta_0|\phi_i\rangle &=& (l^{(i)+}+m_i\alpha^{(i)+}_D)|\lambda_i\rangle, \nonumber \\
\delta_0|C_i\rangle &=& (l^{(i)}_0+m_i^2)|\lambda_i\rangle, \\
\delta_0|D_i\rangle &=& (l^{(i)}+m_i\alpha^{(i)}_D)|\lambda_i\rangle. \nonumber
\end{eqnarray}

Decomposing the equations (\ref{brstmtr}) in terms of the component fields (\ref{massPhi1}), (\ref{massPhi3}) and
(\ref{massLambda1})
 for fields from the first two Hilbert spaces
\begin{eqnarray}\label{ltrm} \nonumber
&&\delta_0 \phi_{\mu_1\mu_2\mu_3}=\partial_{(\mu_1}\lambda_{\mu_2\mu_3)}, \\ \nonumber
&&\delta_0 h_{\mu_1\mu_2}=-\partial_{(\mu_1}\lambda_{\mu_2)}+m\lambda_{\mu_1\mu_2}, \\ \nonumber
&&\delta_0 b_{\mu_1}=\partial_{\mu_1}\lambda +m\lambda_{\mu_1}, \\ \nonumber
&&\delta_0\varphi =m\lambda, \\ \nonumber
&&\delta_0 C_{\mu_1\mu_2} = (\Box -m^2)\lambda_{\mu_1\mu_2}, \\
&&\delta_0 C_{\mu_1} =(\Box -m^2)\lambda_{\mu_1}, \\ \nonumber
&&\delta_0 C = (\Box -m^2)\lambda, \\ \nonumber
&&\delta_0 D_{\mu_1} = \partial_{\mu_2}\lambda_{\mu_1\mu_2}+m\lambda_{\mu_1}, \\ \nonumber
&&\delta_0 D = -\partial_{\mu_1}\lambda_{\mu_1}+2m\lambda, \nonumber
\end{eqnarray}
whereas
for the scalar in the third Hilbert space we have
\be
\delta_0\phi=0.
\ee

The contributions to the gauge transformations at first order in $g$ are given in Appendix \ref{Appendix A}.

\subsection{Equations of motion}

The full Lagrangian is given by \eqref{LIBRSTQ}, where we use the BRST charge \eqref{massbrst}.
The zeroth--order contribution to the Lagrangian is

\begin{eqnarray}
L_0 &=& \sum_{i=1}^2\left[\langle\phi_i|(l_0^{(i)}+m^2)|\phi_i\rangle -\langle\phi_i|(l^{(i)+}+m\alpha^{(i)+}_D)|C_i\rangle
-\langle C_i|(l^{(i)}+m\alpha^{(i)}_D)|\phi_i\rangle
\right. \nonumber \\
&& +\langle C_i||C_i\rangle +\langle C_i|(l^{(i)+}+m\alpha^{(i)+}_D)|D_i\rangle
+\langle D_i|(l^{(i)}+m\alpha^{(i)}_D)|C_i\rangle  \nonumber \\
&& -\langle D_i|(l_0^{(i)}+m^2)|D_i\rangle] 
+\langle\phi_3|(l_0^{(3)}+m_0^2)|\phi_3\rangle.
\end{eqnarray}

The contribution to the first order in $g$
 to the Lagrangian \eqref{LIBRSTQ}, in the case where the cubic interaction vertex is given by $V_6+V_5$, is
\begin{eqnarray}
&& L_1=-\frac{g}{3!3!}\langle\phi_1|\langle\phi_2|\langle\phi_3|(L_\alpha^{(1)+})^3(L_\alpha^{(2)+})^3|0\rangle_{123} \nonumber \\
&&
-\frac{ga_1^2}{4}\langle C_1|\langle C_2|\langle\phi_3|(L_\alpha^{(1)+})^2(L_\alpha^{(2)+})^2|0\rangle_{123} \nonumber \\
&& -\frac{g}{4}\langle\phi_1|\langle\phi_2|\langle\phi_3|(L_\alpha^{(1)+})^2(L_\alpha^{(2)+})^2\hat{Q}_\alpha^{(12)+}|0\rangle_{123}
 \nonumber \\
&& -\frac{ga_2}{8ma_1}\langle\phi_1|\langle\phi_2|\langle\phi_3|\left[\alpha^{(1)+}_D L_\alpha^{(2)+}-\alpha^{(2)+}_D L_\alpha^{(1)+}\right](L_\alpha^{(1)+})^2 (L_\alpha^{(2)+})^2|0\rangle_{123}
\nonumber \\
&& -ga_1^2\langle C_1|\langle C_2|\langle\phi_3| L_\alpha^{(1)+}L_\alpha^{(2)+}\hat{Q}_\alpha^{(12)+}|0\rangle_{123}
\nonumber \\
&& -\frac{3ga_1a_2}{4m}\langle C_1|\langle C_2|\langle\phi_3|\left[\alpha^{(1)+}_D L_\alpha^{(2)+}-\alpha^{(2)+}_D L_\alpha^{(1)+}\right] L_\alpha^{(1)+} L_\alpha^{(2)+}|0\rangle_{123}  \\
&& -\frac{ga_1a_2}{4}\langle C_1|\langle D_2|\langle\phi_3|(L_\alpha^{(1)+})^2 L_\alpha^{(2)+}|0\rangle_{123}+\frac{ga_1a_2}{4}\langle D_1|\langle C_2|\langle\phi_3|L_\alpha^{(1)+} (L_\alpha^{(2)+})^2|0\rangle_{123} \nonumber \\
&& +h.c., \nonumber
\end{eqnarray}
where $L^{(i)+}_\alpha$ is the part of \eqref{f-1}  containing only oscillators, whilst $\hat{Q}^{(12)+}_\alpha$ is given in \eqref{qhatdef}.

Therefore the equations of motion read:

With respect to $\langle\phi_1|$
\begin{eqnarray}
&& (l_0^{(1)}+m^2)|\phi_1\rangle =(l^{(1)+}+m\alpha^{(1)+}_D)|C_1\rangle +\frac{g}{3!3!}\langle\phi_2|\langle\phi_3|(L_\alpha^{(1)})^3(L_\alpha^{(2)})^3|0\rangle_{123} \nonumber \\
&& +\frac{g}{4}\langle\phi_2|\langle\phi_3|(L_\alpha^{(1)+})^2(L_\alpha^{(2)+})^2\hat{Q}_\alpha^{(12)+}|0\rangle_{123} \nonumber \\
&& +\frac{ga_2}{8ma_1}\langle\phi_2|\langle\phi_3|\left[\alpha^{(1)+}_D L_\alpha^{(2)+}-\alpha^{(2)+}_D L_\alpha^{(1)+}\right](L_\alpha^{(1)+})^2 (L_\alpha^{(2)+})^2|0\rangle_{123}.
\end{eqnarray}

With respect to $\langle C_1|$:

\begin{eqnarray}
&&|C_1\rangle =(l^{(1)}+m\alpha^{(1)}_D)|\phi_1\rangle -(l^{(1)+}\! +m\alpha^{(1)+}_D)|D_1\rangle +\frac{ga_1^2}{4}\langle C_2|\langle\phi_3|(L_\alpha^{(1)})^2(L_\alpha^{(2)})^2|0\rangle_{123} \nonumber \\
&& +ga_1^2\langle C_2|\langle\phi_3| L_\alpha^{(1)+}L_\alpha^{(2)+}\hat{Q}_\alpha^{(12)+}|0\rangle_{123} +\frac{3ga_1a_2}{4m}\langle C_2|\langle\phi_3|\alpha^{(1)+}_D L_\alpha^{(1)+}(L_\alpha^{(2)+})^2|0\rangle_{123} \nonumber \\
&& -\frac{3ga_1a_2}{4m}\langle C_2|\langle\phi_3|\alpha^{(2)+}_D (L_\alpha^{(1)+})^2 L_\alpha^{(2)+}|0\rangle_{123}
+\frac{ga_1a_2}{4}\langle D_2|\langle\phi_3|(L_\alpha^{(1)+})^2 L_\alpha^{(2)+}|0\rangle_{123}. \nonumber \\
\end{eqnarray}

With respect to $\langle D_1|$:

\begin{equation}
(l_0^{(1)}+m^2)|D_1\rangle =(l^{(1)}+m\alpha^{(1)}_D)|C_1\rangle +\frac{ga_1a_2}{4}\langle C_2|\langle\phi_3|L_\alpha^{(1)+} (L_\alpha^{(2)+})^2|0\rangle_{123}.
\end{equation}

Let us first consider the equation of motion for $\phi_{\mu_1\mu_2\mu_3}$, which gives

\begin{eqnarray}\label{massphi3}
&& (\Box -m^2)\phi_{\mu_1\mu_2\mu_3} = \partial_{(\mu_1}C_{\mu_2\mu_3)} \nonumber \\
&& +\frac{ga_1^6}{3!}\left[8(\partial_{\mu_1\mu_2\mu_3}\phi_{\nu_1\nu_2\nu_3})(\partial_{\nu_1\nu_2\nu_3}\phi)+12(\partial_{\mu_1\mu_2\mu_3\nu_1}\phi_{\nu_1\nu_2\nu_3})(\partial_{\nu_2\nu_3}\phi) \right. \nonumber \\
&& +6(\partial_{\mu_1\mu_2\mu_3\nu_1\nu_2}\phi_{\nu_1\nu_2\nu_3})(\partial_{\nu_3}\phi)
 +(\partial_{\mu_1\mu_2\mu_3\nu_1\nu_2\nu_3}\phi_{\nu_1\nu_2\nu_3})\phi \nonumber \\
 && -24(\partial_{\mu_1\mu_2}\phi_{\nu_1\nu_2\nu_3})(\partial_{\mu_3\nu_1\nu_2\nu_3}\phi)-36(\partial_{\mu_1\mu_2\nu_1}\phi_{\nu_1\nu_2\nu_3})(\partial_{\mu_3\nu_2\nu_3}\phi) \nonumber \\
&& -18(\partial_{\mu_1\mu_2\nu_1\nu_2}\phi_{\nu_1\nu_2\nu_3})(\partial_{\mu_3\nu_3}\phi)-3(\partial_{\mu_1\mu_2\nu_1\nu_2\nu_3}\phi_{\nu_1\nu_2\nu_3})(\partial_{\mu_3}\phi) \nonumber \\
&& +24(\partial_{\mu_1}\phi_{\nu_1\nu_2\nu_3})(\partial_{\mu_2\mu_3\nu_1\nu_2\nu_3}\phi) +36(\partial_{\mu_1\nu_1}\phi_{\nu_1\nu_2\nu_3})(\partial_{\mu_2\mu_3\nu_2\nu_3}\phi) \nonumber \\
&& +18(\partial_{\mu_1\nu_1\nu_2}\phi_{\nu_1\nu_2\nu_3})(\partial_{\mu_2\mu_3\nu_3}\phi)
+3(\partial_{\mu_1\nu_1\nu_2\nu_3}\phi_{\nu_1\nu_2\nu_3})(\partial_{\mu_2\mu_3}\phi) \nonumber \\
&& -8\phi_{\nu_1\nu_2\nu_3}(\partial_{\mu_1\mu_2\mu_3\nu_1\nu_2\nu_3}\phi)-12(\partial_{\nu_1}\phi_{\nu_1\nu_2\nu_3})(\partial_{\mu_1\mu_2\mu_3\nu_2\nu_3}\phi) \nonumber \\
&& \left. -6(\partial_{\nu_1\nu_2}\phi_{\nu_1\nu_2\nu_3})(\partial_{\mu_1\mu_2\mu_3\nu_3}\phi) -(\partial_{\nu_1\nu_2\nu_3}\phi_{\nu_1\nu_2\nu_3})(\partial_{\mu_1\mu_2\mu_3}\phi)\right] \\
&& +\frac{3ga_1^4a_2}{2}\left[4(\partial_{\mu_1\mu_2}\phi_{\nu_1\nu_2\mu_3})(\partial_{\nu_1\nu_2}\phi)+4(\partial_{\mu_1\mu_2\nu_1}\phi_{\nu_1\nu_2\mu_3})(\partial_{\nu_2}\phi)+(\partial_{\mu_1\mu_2\nu_1\nu_2}\phi_{\nu_1\nu_2\mu_3})\phi \right. \nonumber \\
&& -8(\partial_{\mu_1}\phi_{\nu_1\nu_2\mu_3})(\partial_{\mu_2\nu_1\nu_2}\phi)-8(\partial_{\mu_1\nu_1}\phi_{\nu_1\nu_2\mu_3})(\partial_{\mu_2\nu_2}\phi) -2(\partial_{\mu_1\nu_1\nu_2}\phi_{\nu_1\nu_2\mu_3})(\partial_{\mu_2}\phi) \nonumber \\
&& \left.+4\phi_{\nu_1\nu_2\mu_3}(\partial_{\mu_1\mu_2\nu_1\nu_2}\phi)+4(\partial_{\nu_1}\phi_{\nu_1\nu_2\mu_3})(\partial_{\mu_1\mu_2\nu_2}\phi)+(\partial_{\nu_1\nu_2}\phi_{\nu_1\nu_2\mu_3})(\partial_{\mu_1\mu_2}\phi)\right] \nonumber \\
&& -\frac{3ga_1^4a_2}{4m}
\left[4(\partial_{\mu_1\mu_2\mu_3}h_{\nu_1\nu_2})(\partial_{\nu_1\nu_2}\phi)+4(\partial_{\mu_1\mu_2\mu_3\nu_1}h_{\nu_1\nu_2})(\partial_{\nu_2}\phi)+(\partial_{\mu_1\mu_2\mu_3\nu_1\nu_2}h_{\nu_1\nu_2})\phi
\right. \nonumber \\
&& -12(\partial_{\mu_1\mu_2}h_{\nu_1\nu_2})(\partial_{\mu_3\nu_1\nu_2}\phi)-12(\partial_{\mu_1\mu_2\nu_1}h_{\nu_1\nu_2})(\partial_{\mu_3\nu_2}\phi)-3(\partial_{\mu_1\mu_2\nu_1\nu_2}h_{\nu_1\nu_2})(\partial_{\mu_3}\phi) \nonumber \\
&& +12(\partial_{\mu_1}h_{\nu_1\nu_2})(\partial_{\mu_2\mu_3\nu_1\nu_2}\phi)+12(\partial_{\mu_1\nu_1}h_{\nu_1\nu_2})(\partial_{\mu_2\mu_3\nu_2}\phi)+3(\partial_{\mu_1\nu_1\nu_2}h_{\nu_1\nu_2})(\partial_{\mu_2\mu_3}\phi) \nonumber \\
&& \left.-4 h_{\nu_1\nu_2}(\partial_{\mu_1\mu_2\mu_3\nu_1\nu_2}\phi)-4(\partial_{\nu_1}h_{\nu_1\nu_2})(\partial_{\mu_1\mu_2\mu_3\nu_2}\phi)-(\partial_{\nu_1\nu_2}h_{\nu_1\nu_2})(\partial_{\mu_1\mu_2\mu_3}\phi)
\right]. \nonumber
\end{eqnarray}

The equation of motion for $h_{\mu_1\mu_2}$ is

\begin{eqnarray}\label{massh2}
&& (\Box -m^2)h_{\mu_1\mu_2} =-\partial_{(\mu_1}C_{\mu_2)}+mC_{\mu_1\mu_2} \nonumber \\
&& +\frac{ga_1^4a_2}{4}\left[4(\partial_{\mu_1\mu_2}h_{\nu_1\nu_2})(\partial_{\nu_1\nu_2}\phi)+4(\partial_{\mu_1\mu_2\nu_1}h_{\nu_1\nu_2})(\partial_{\nu_2}\phi)+(\partial_{\mu_1\mu_2\nu_1\nu_2}h_{\nu_1\nu_2})\phi \right. \nonumber \\
&& -8(\partial_{\mu_1}h_{\nu_1\nu_2})(\partial_{\mu_2\nu_1\nu_2}\phi)-8(\partial_{\mu_1\nu_1}h_{\nu_1\nu_2})(\partial_{\mu_2\nu_2}\phi) -2(\partial_{\mu_1\nu_1\nu_2}h_{\nu_1\nu_2})(\partial_{\mu_2}\phi) \nonumber \\
&& \left.+4 h_{\nu_1\nu_2}(\partial_{\mu_1\mu_2\nu_1\nu_2}\phi)+4(\partial_{\nu_1}h_{\nu_1\nu_2})(\partial_{\mu_1\mu_2\nu_2}\phi)+(\partial_{\nu_1\nu_2}h_{\nu_1\nu_2})(\partial_{\mu_1\mu_2}\phi)\right] \nonumber \\
&& +\frac{ga_1^4a_2}{4m}\left[8(\partial_{\mu_1\mu_2}\phi_{\nu_1\nu_2\nu_3})(\partial_{\nu_1\nu_2\nu_3}\phi)+12(\partial_{\mu_1\mu_2\nu_1}\phi_{\nu_1\nu_2\nu_3})(\partial_{\nu_2\nu_3}\phi) \right.  \\
&& +6(\partial_{\mu_1\mu_2\nu_1\nu_2}\phi_{\nu_1\nu_2\nu_3})(\partial_{\nu_3}\phi)
 +(\partial_{\mu_1\mu_2\nu_1\nu_2\nu_3}\phi_{\nu_1\nu_2\nu_3})\phi \nonumber \\
 &&-16(\partial_{\mu_1}\phi_{\nu_1\nu_2\nu_3})(\partial_{\mu_2\nu_1\nu_2\nu_3}\phi)-24(\partial_{\mu_1\nu_1}\phi_{\nu_1\nu_2\nu_3})(\partial_{\mu_2\nu_2\nu_3}\phi) \nonumber \\
&& -12(\partial_{\mu_1\nu_1\nu_2}\phi_{\nu_1\nu_2\nu_3})(\partial_{\mu_2\nu_3}\phi)-2(\partial_{\mu_1\nu_1\nu_2\nu_3}\phi_{\nu_1\nu_2\nu_3})(\partial_{\mu_2}\phi) +8\phi_{\nu_1\nu_2\nu_3}(\partial_{\mu_1\mu_2\nu_1\nu_2\nu_3}\phi) \nonumber \\
&& \left. +12(\partial_{\nu_1}\phi_{\nu_1\nu_2\nu_3})(\partial_{\mu_1\mu_2\nu_2\nu_3}\phi)+6(\partial_{\nu_1\nu_2}\phi_{\nu_1\nu_2\nu_3})(\partial_{\mu_1\mu_2\nu_3}\phi)+(\partial_{\nu_1\nu_2\nu_3}\phi_{\nu_1\nu_2\nu_3})(\partial_{\mu_1\mu_2}\phi)\right]. \nonumber
\end{eqnarray}

The equation of motion for $b_{\mu_1}$ is

\begin{equation}\label{massb1}
(\Box -m^2)b_{\mu_1} =\partial_{\mu_1}C+mC_{\mu_1}.
\end{equation}

The equation of motion for $\varphi$ is

\begin{equation}\label{massvarphi0}
 (\Box -m^2)\varphi =mC.
\end{equation}

The equation of motion for $C_{\mu_1\mu_2}$ is

\begin{eqnarray}\label{massC2}
&& C_{\mu_1\mu_2} =\partial_{\mu_3}\phi_{\mu_1\mu_2\mu_3}-mh_{\mu_1\mu_2}-\partial_{(\mu_1}D_{\mu_2)} \nonumber \\
&& -\frac{ga_1^6}{2!}\left[4(\partial_{\mu_1\mu_2}C_{\nu_1\nu_2})(\partial_{\nu_1\nu_2}\phi)+4(\partial_{\mu_1\mu_2\nu_1}C_{\nu_1\nu_2})(\partial_{\nu_2}\phi)+(\partial_{\mu_1\mu_2\nu_1\nu_2}C_{\nu_1\nu_2})\phi
\right. \nonumber \\
&&-8(\partial_{\mu_1}C_{\nu_1\nu_2})(\partial_{\mu_2\nu_1\nu_2}\phi)-8(\partial_{\mu_1\nu_1}C_{\nu_1\nu_2})(\partial_{\mu_2\nu_2}\phi)-2(\partial_{\mu_1\nu_1\nu_2}C_{\nu_1\nu_2})(\partial_{\mu_2}\phi) \nonumber \\
&& \left. +4C_{\nu_1\nu_2}(\partial_{\mu_1\mu_2\nu_1\nu_2}\phi)+4(\partial_{\nu_1}C_{\nu_1\nu_2})(\partial_{\mu_1\mu_2\nu_2}\phi)+(\partial_{\nu_1\nu_2}C_{\nu_1\nu_2})(\partial_{\mu_1\mu_2}\phi)\right] \nonumber \\
&& +2ga_1^4a_2\left[2(\partial_{\mu_1}C_{\nu_1\mu_2})(\partial_{\nu_1}\phi)+(\partial_{\mu_1\nu_1}C_{\nu_1\mu_2})\phi -2C_{\nu_1\mu_2}(\partial_{\mu_1\nu_1}\phi)-(\partial_{\nu_1}C_{\nu_1\mu_2})(\partial_{\mu_1}\phi)
\right] \nonumber \\
&& +\frac{3ga_1^4a_2}{2m}\left[2(\partial_{\mu_1\mu_2}C_{\nu_1})(\partial_{\nu_1}\phi)+(\partial_{\mu_1\mu_2\nu_1}C_{\nu_1})\phi -4(\partial_{\mu_1}C_{\nu_1})(\partial_{\mu_2\nu_1}\phi)
\right. \nonumber \\
&& \left. -2(\partial_{\mu_1\nu_1}C_{\nu_1})(\partial_{\mu_2}\phi)+2C_{\nu_1}(\partial_{\mu_1\mu_2\nu_1}\phi)+(\partial_{\nu_1}C_{\nu_1})(\partial_{\mu_1\mu_2}\phi)\right] \nonumber \\
&& +\frac{ga_1^4a_2}{2}\left[2(\partial_{\mu_1\mu_2}D_{\nu_1})(\partial_{\nu_1}\phi)+(\partial_{\mu_1\mu_2\nu_1}D_{\nu_1})\phi -4(\partial_{\mu_1}D_{\nu_1})(\partial_{\mu_2\nu_1}\phi)
\right. \nonumber \\
&& \left. -2(\partial_{\mu_1\nu_1}D_{\nu_1})(\partial_{\mu_2}\phi)+2D_{\nu_1}(\partial_{\mu_1\mu_2\nu_1}\phi)+(\partial_{\nu_1}D_{\nu_1})(\partial_{\mu_1\mu_2}\phi)\right].
\end{eqnarray}

The equation of motion for $C_{\mu_1}$ is

\begin{eqnarray}\label{massC1}
&& C_{\mu_1} =-\partial_{\mu_2}h_{\mu_1\mu_2}-2mb_{\mu_1}+\partial_{\mu_1}D+mD_{\mu_1} \nonumber \\
&& +\frac{ga_1^4a_2}{2}\left[2(\partial_{\mu_1}C_{\nu_1})(\partial_{\nu_1}\phi)+(\partial_{\mu_1\nu_1}C_{\nu_1})\phi -2C_{\nu_1}(\partial_{\mu_1\nu_1}\phi)-(\partial_{\nu_1}C_{\nu_1})(\partial_{\mu_1}\phi)\right]. \nonumber \\
\end{eqnarray}

The equation of motion for $C$ is

\begin{equation}\label{massC0}
 C=\partial_{\mu_1}b_{\mu_1}-3m\varphi +mD.
\end{equation}

The equation of motion for $D_{\mu_1}$ is

\begin{eqnarray}\label{massD1}
&& (\Box -m^2)D_{\mu_1}=\partial_{\mu_2}C_{\mu_1\mu_2}+mC_{\mu_1} \nonumber \\
&& -\frac{ga_1^4a_2}{4}\left[4(\partial_{\mu_1}C_{\nu_1\nu_2})(\partial_{\nu_1\nu_2}\phi)+4(\partial_{\mu_1\nu_1}C_{\nu_1\nu_2})(\partial_{\nu_2}\phi)+(\partial_{\mu_1\nu_1\nu_2}C_{\nu_1\nu_2})\phi
\right. \nonumber \\
&& \left. -4C_{\nu_1\nu_2}(\partial_{\mu_1\nu_1\nu_2}\phi)-4(\partial_{\nu_1}C_{\nu_1\nu_2})(\partial_{\mu_1\nu_2}\phi)-(\partial_{\nu_1\nu_2}C_{\nu_1\nu_2})(\partial_{\mu_1}\phi)\right].
\end{eqnarray}

The equation of motion for $D$ is

\begin{equation}\label{massD0}
(\Box -m^2)D =-\partial_{\mu_1}C_{\mu_1}+2mC.
\end{equation}

\subsection{Gauge fixing}\label{3-3-0gf}

In analogy with the case of massless fields considered in  Section \ref{gfpm}
let us discuss first the gauge fixing procedure in the absence of interactions \cite{Pashnev:1997rm}.

As one can see from the equation (\ref{ltrm})
one can use the parameter $\lambda$ to gauge away the field $\varphi$, then one can use the parameter $\lambda_\mu$
to gauge away the field $b_\mu$ and finally use the parameter $\lambda_{\mu_1 \mu_2}$ to gauge away the
field $h_{\mu_1 \mu_2}$.

Further, the equation \eqref{massvarphi0} imposes $C=0$, the equation  \eqref{massb1} imposes $C_{\mu}=0$,
and the equation \eqref{massC2} imposes $C_{\mu_1 \mu_2}=0$.

 We can then use \eqref{massC0} to show $D=0$ and \eqref{massC1} to show $D_{\mu}=0$. With this choice, the equation of motion \eqref{massD0} for $D$ is trivially satisfied.

Therefore one ends up with the field satisfying a mass-shell
\be
 (\Box -m^2)\phi_{\mu_1\mu_2\mu_3} =0,
\ee
and transversality \eqref{tr-m} conditions.

Now let us discus what happens in the presence of interactions, in particular when we consider the cubic vertex
which consists of parts $V_6$ and $V_5$ given in \eqref{V6V5}.
 If one considers {\it only} $V_6$, then the procedure outlined above
remains unmodified, and after the gauge fixing
one obtains only the physical field $\phi_{\mu_1 \mu_2 \mu_3}$
which satisfies the
transversality \eqref{tr-m} condition and a nonlinear equation of motion \eqref{massphi3} with the parameter
$a_2$ set equal to zero.
Therefore in this case the degrees of freedom are the same as for the free field.

If one considers {\it both} $V_6$ and $V_5$ then the situation is the following.
After the gauge fixing, i.e. elimination of the
St\"{u}ckelberg fields $h_{\mu_1 \mu_2}$, $b_\mu$ and $\varphi$ via gauge transformations,
the equations of motion put the fields $C_{\mu}, C, D_\mu$ and $D$ to be equal to zero.
However the field $C_{\mu_1 \mu_2}$ which represents longitudinal (nonphysical)
modes of the field $\phi_{\mu_1 \mu_2 \mu_3}$ is no longer zero, but rather it satisfies the equation

\begin{eqnarray}\label{massh2gf}
&& 0=mC_{\mu_1\mu_2} \nonumber \\
&& +\frac{ga_1^4a_2}{4m}
\left[8(\partial_{\mu_1\mu_2}\phi_{\nu_1\nu_2\nu_3})(\partial_{\nu_1\nu_2\nu_3}\phi)+12(\partial_{\mu_1\mu_2\nu_1}\phi_{\nu_1\nu_2\nu_3})(\partial_{\nu_2\nu_3}\phi) \right. \nonumber \\
&& +6(\partial_{\mu_1\mu_2\nu_1\nu_2}\phi_{\nu_1\nu_2\nu_3})(\partial_{\nu_3}\phi) +(\partial_{\mu_1\mu_2\nu_1\nu_2\nu_3}\phi_{\nu_1\nu_2\nu_3})\phi  -16(\partial_{\mu_1}\phi_{\nu_1\nu_2\nu_3})(\partial_{\mu_2\nu_1\nu_2\nu_3}\phi) \nonumber \\
&&-24(\partial_{\mu_1\nu_1}\phi_{\nu_1\nu_2\nu_3})(\partial_{\mu_2\nu_2\nu_3}\phi) -12(\partial_{\mu_1\nu_1\nu_2}\phi_{\nu_1\nu_2\nu_3})(\partial_{\mu_2\nu_3}\phi)\nonumber \\
&& -2(\partial_{\mu_1\nu_1\nu_2\nu_3}\phi_{\nu_1\nu_2\nu_3})(\partial_{\mu_2}\phi) +8\phi_{\nu_1\nu_2\nu_3}(\partial_{\mu_1\mu_2\nu_1\nu_2\nu_3}\phi)+12(\partial_{\nu_1}\phi_{\nu_1\nu_2\nu_3})(\partial_{\mu_1\mu_2\nu_2\nu_3}\phi) \nonumber \\
&& \left. +6(\partial_{\nu_1\nu_2}\phi_{\nu_1\nu_2\nu_3})(\partial_{\mu_1\mu_2\nu_3}\phi)+(\partial_{\nu_1\nu_2\nu_3}\phi_{\nu_1\nu_2\nu_3})(\partial_{\mu_1\mu_2}\phi)\right], \nonumber \\
\end{eqnarray}
which is the equation of motion with respect to the St\"{u}ckelberg field $h_{\mu_1\mu_2}$.

Let us consider now the scalar as a background field. As one can see
from (\ref{massh2gf})
 one can consistently impose the transversality condition
provided the following constraint on the background is satisfied
\begin{eqnarray}\label{gfcondition}
&& 0= g\frac{a_1^4a_2}{4m}\left[8(\partial_{\mu_1\mu_2}\phi_{\nu_1\nu_2\nu_3})(\partial_{\nu_1\nu_2\nu_3}\phi)-16(\partial_{\mu_1}\phi_{\nu_1\nu_2\nu_3})(\partial_{\mu_2\nu_1\nu_2\nu_3}\phi)\right. \nonumber \\
&& \left. \hspace{10mm}+8\phi_{\nu_1\nu_2\nu_3}(\partial_{\mu_1\mu_2\nu_1\nu_2\nu_3}\phi)\right].
\end{eqnarray}
This constraint in turn means that one of the two conditions should be satisfied
\begin{itemize}

\item The triple and higher derivatives on the background field are zero.

\item The mass parameter is large enough that one can ignore the term in the action \eqref{LIBRSTQ}
 which gives rise to the right--hand
side of the equation \eqref{gfcondition}.

\end{itemize}

We shall see that after the Velo-Zwanziger like analysis of
causality the first option will be ruled out.

Let us summarize the results of this section.
For the case of massive higher spin fields one again starts from the gauge invariant free Lagrangian
which unlike the one for massless  higher spin fields contains also
 St\"{u}ckelberg fields. The number of the parameters of gauge transformations
precisely equals  the number of the  St\"{u}ckelberg fields. After
completely using  the gauge freedom to eliminate the
St\"{u}ckelberg fields, the other auxiliary degrees of freedom are
eliminated and the transversality condition on the physical field is
imposed due to the free equations of motion. When performing the
nonlinear (cubic) deformation of the system the number of  gauge
transformation parameters is preserved which again allows one to
gauge away  all St\"{u}ckelberg fields. However since the free
equations of motion are modified, then, unlike the case of massless
fields, for the massive ones the transversality condition can be
modified as well. As a result, the gauge invariance is not enough to
determine the cubic vertex, and to preserve the true number of
degrees of freedom compatible with the transversality condition one
needs to impose  additional restrictions on the free parameters
of the interaction vertex.

\setcounter{equation}0
\section{Causality Analysis for 3-3-0 System}\label{3-3-0causality}

In this  section we perform an analysis of causality for this system. In
the case under consideration we have a coupled system of equations
of motion for spin $3$ and spin $0$ fields. The features of this
system are the higher derivatives in the interaction terms. It is
clear that this system is incomplete, since  one can expect an
infinite system of equations involving fields of all higher spins
with an infinite number of interaction vertices where the vertices can
include an arbitrary number of derivatives. Any truncation will be only
approximate. Therefore, it is not completely clear how causality
analysis for the higher spin field theory can be carried out at all.
In contrast with String Theory, which can be treated as a higher spin
field model, where causality of the system of higher spin field
equations of motion is stipulated by the underlying fundamental causal
string equations of motion, in the higher spin field theory such an
underlying theory is unknown. Nevertheless we will try to develop the
causality analysis for our truncated theory in the framework of some
simplified model.

Our setting is as follows. First,  we consider the scalar as a
background field satisfying the free equations of motion
\begin{equation}\label{zo}
(l_0^{(3)}-m^2)|\phi_3\rangle +\mathcal{O}(g)=0,
\end{equation}
i.e. $(\Box-m^2)\phi=0$. This equation can be derived from the
following consideration. Let us consider the coupled system of
equations for spin 3 and spin 0 fields. The solution to the scalar field
equation is the free field satisfying (\ref{zo}) plus the terms
 $\mathcal{O}(g)$. If we substitute such a solution into the equation for the spin $3$
field we will see that the terms $\mathcal{O}(g)$ in the solution for scalar field
can be omitted.

Second, we  take  the some kind of low-energy approximation (i.e.
the case of small $p_\mu$) where we can ignore the third derivative
acting on a dynamical field in comparison to the second one. Using
the results from the previous section we obtain the following
equation for the field $\phi_{\mu_1 \mu_2 \mu_3}$
\begin{eqnarray}\label{imp}
&& (\Box - m^2)\phi_{\mu_1\mu_2\mu_3} +\frac{ga_1^6}{3!}\left[
24(\partial_{\mu_1\mu_2}\phi_{\nu_1\nu_2\nu_3})(\partial_{\mu_3\nu_1\nu_2\nu_3}\phi)
\right. \nonumber \\
&& \left. -24(\partial_{\mu_1}\phi_{\nu_1\nu_2\nu_3})(\partial_{\mu_2\mu_3\nu_1\nu_2\nu_3}\phi)+8\phi_{\nu_1\nu_2\nu_3}(\partial_{\mu_1\mu_2\mu_3\nu_1\nu_2\nu_3}\phi)\right] \nonumber \\
&& +\frac{3ga_1^4 a_2}{2}\left[ -4(\partial_{\mu_1\mu_2}\phi_{\nu_1\nu_2\mu_3})(\partial_{\nu_1\nu_2}\phi)+8(\partial_{\mu_1}\phi_{\nu_1\nu_2\mu_3})(\partial_{\mu_2\nu_1\nu_2}\phi)\right. \nonumber \\
&& \left. -4\phi_{\nu_1\nu_2\mu_3}(\partial_{\mu_1\mu_2\nu_1\nu_2}\phi)
\right]=0,
\end{eqnarray}
along with the transversality condition (\ref{tr-m}).
As the first step  we shall  perform an analysis of the causality without taking into account the condition on the background
(\ref{gfcondition}). Then as the second step we consider the implications of the equation (\ref{gfcondition}).

In order to perform the causality analysis in a manner analogous to \cite{Velo:1969bt} we consider the
 terms  in (\ref{imp}) with two derivatives acting on the dynamical field. Let us denote
 derivatives of the background field $\phi$ by
\[
G_{\rho_1\ldots\rho_k}:=\partial_{\rho_1\ldots\rho_k}\phi,
\]
 and define

\begin{equation}\label{Gtilde}
\tilde{G}_{\rho_1\ldots\rho_k}=p_{\rho_1}\ldots p_{\rho_k} \phi(p).
\end{equation}
Note that we have removed the factors of $i$ in relation to the usual Fourier transform of $G_{\rho_1\ldots\rho_k}$.

Hence, we are led to consider the object
\begin{equation}\label{vzmatrix}
\left[-p^2\delta^{\nu_1}_{(\mu_1}\delta^{\nu_2}_{\mu_2}\delta^{\nu_3}_{\mu_3)}+
4ga_1^6 p_{(\mu_1}p_{\mu_2}{\tilde G}_{\mu_3)}{}^{\nu_1\nu_2\nu_3}-
6ga_1^4 a_2 p_{(\mu_1}p_{\mu_2}\delta^{\nu_3}_{\mu_3)}
{\tilde G}^{\nu_1\nu_2}\right]\phi_{\nu_1\nu_2\nu_3},
\end{equation}
which we can think of as a $N\times N$ matrix, with $N=\binom{\mathcal{D}+2}{3}$
\footnote{Recall that in general the number of
 independent components of a rank-$s$ totally symmetric tensor in $\mathcal{D}$ dimensions is $\binom{\mathcal{D}-1+s}{s}$.},
 acting on the space of totally-symmetric 3-tensors which we take to have the basis

\[
\left\lbrace\phi_{000},\phi_{00i},\phi_{0ii},\phi_{0ij},\phi_{iii},\phi_{iij},\phi_{ijj},\phi_{ijk}\right\rbrace,
\]
for $i=1,\ldots,\mathcal{D}-1$ and $i<j<k$.

For example, the ``1-1'' component of \eqref{vzmatrix} will be

\[
-p^2+4ga_1^6(p^0)^2{\tilde G}_0^{000} -6ga_1^4 a_2(p^0)^2 {\tilde G}^{00}.
\]

In order to carry out the causality analysis \textit{\`{a} la} Velo-Zwanziger we need
 to calculate the determinant of this matrix. Before we do so however, let us make some assumptions. In particular, we take the coupling constant $g$ to be small, i.e. we ignore all terms of $\mathcal{O}(g^2)$, which is reasonable given that we are working in a perturbative framework.

Then the determinant of the matrix in \eqref{vzmatrix} can be written as

\begin{equation}\label{vzdet1}
D(p)=\det(-p^2\mathbb{I}_{N}+A)=(-p^2)^{N-1}\left[-(p^2)+\textrm{tr}(A)+\mathcal{O}(g^2)\right],
\end{equation}
where $A$ only contains terms proportional to the background field (and hence $g$), and we have ignored higher-order contributions.
In particular, we have
\begin{eqnarray}
A_{(\mu_1\mu_2\mu_3)}^{(\nu_1\nu_2\nu_3)} &=& \frac{4}{3}ga_1^6\left[p_{\mu_1}p_{\mu_2}
{\tilde G}_{\mu_3}^{\nu_1\nu_2\nu_3}+
p_{\mu_1}p_{\mu_3} {\tilde G}_{\mu_2}^{\nu_1\nu_2\nu_3}+p_{\mu_2}p_{\mu_3}
{\tilde G}_{\mu_1}^{\nu_1\nu_2\nu_3}\right] \nonumber \\
&& -2ga_1^4 a_2\left[p_{\mu_1}p_{\mu_2}\delta_{\mu_3}^{\nu_3}
{\tilde G}^{\nu_1\nu_2}+
p_{\mu_1}p_{\mu_3}\delta_{\mu_2}^{\nu_3}{\tilde G}^{\nu_1\nu_2}
+p_{\mu_2}p_{\mu_3}\delta_{\mu_1}^{\nu_3}{\tilde G}^{\nu_1\nu_2}\right]. \nonumber \\
\end{eqnarray}
Finding the trace of this matrix then just amounts to calculating $A_{(\mu_1\mu_2\mu_3)}^{(\mu_1\mu_2\mu_3)}$, which is
\begin{eqnarray}\label{trA-1}
\textrm{tr}(A)&=&4ga_1^6 p_{\mu_1}p_{\mu_2}{\tilde G}_{\mu_3}^{\mu_1\mu_2\mu_3} -2ga_1^4a_2(\mathcal{D}+2)p_{\mu_1}p_{\mu_2}{\tilde G}^{\mu_1\mu_2}  \\ \nonumber
&=& 2ga_1^4 p_{\mu_1}p_{\mu_2}\left(2a_1^2 p^2 -(\mathcal{D}+2)a_2\right){\tilde G}^{\mu_1\mu_2}.
\end{eqnarray}
Hence, the determinant \eqref{vzdet1} is given by
\begin{equation}\label{vzdet2}
D(p)=(p^2)^{N-1}\left[p^2+2ga_1^4p_{\mu_1}p_{\mu_2}\left(2a_1^2 p^2-(\mathcal{D}+2)a_2\right){\tilde G}^{\mu_1\mu_2}\right].
\end{equation}
Assuming that the background scalar satisfies
the zeroth--order equations of motion (\ref{zo})
 and setting a condition on the parameters $a_1$ and $a_2$
\begin{equation}\label{spin3caus}
a_2=-\frac{2}{\mathcal{D}+2}a_1^2m^2,
\end{equation}
then \eqref{vzdet2} reduces to  $D(p)=(p^2)^{N}$, and so we have causal propagation.
On the other hand, if the equation (\ref{spin3caus}) is not satisfied the causality is broken.

However
after we performed the analysis of causality for the system
 we  also   have to take into account the condition of preservation of degrees of freedom
discussed in the previous Section.

The discussion goes as follows. First
recall that if we take only the $V_6$ part of the vertex (\ref{V6V5})
 then the number of degrees of freedom in the spin $3$
triplet is unchanged without imposing any extra condition on the background.
However choosing a vertex which contains only $V_6$ implies that the constant
$a_2$ is zero. Therefore due to (\ref{spin3caus}) the other constant $a_1$
is zero as well and so one has no interaction.
This is in contradiction with our initial assumption that $a_1$ is non-zero.
Therefore in this case the condition of correct degrees of freedom for the higher spin field and the condition of
the causal propagation are not compatible with each other.

If one takes both the $V_6$ and $V_5$ parts in the vertex, then as we saw in the previous section
one also needs to take into account the condition (\ref{gfcondition}).
Then, as we have seen, one has two options to solve it. If one requires that
the triple and higher derivatives on the background scalar vanish,
then the equation  (\ref{trA-1}) implies that
$a_1^4 a_2=0$ which is in contradiction with the original  assumption
that both $a_1, a_2 \neq 0$.
Therefore as in the previous case we have no interaction or in other words the cubic interaction
 allows no causal propagation for the massive fields with spin $3$.

The second possibility for the preservation of degrees of freedom,
i.e. the situation when the right--hand side of the equation
(\ref{gfcondition}) vanishes due to the large mass
parameter\footnote{More precisely one requires that the terms
suppressed  by the  factor $\frac{1}{m}$ in the action can be
ignored in comparison with the terms that give us the equations of
motion (\ref{imp}). This condition also implies that the coupling
constants and the derivatives of the background fields are of order
of one or smaller.} in turn does not impose any further restriction
on the parameters $a_1$ and $a_2$ apart from the one
(\ref{spin3caus}) which results from the causality analysis.
Therefore one obtains that in this particular case one has causal
propagation of the massive spin three field, coupled to a background
scalar.

A conclusion which can be drawn from the gauge fixing procedure and from the
causality analysis for  massive higher spin fields is that the gauge invariance  and
the presence of correct number degrees of freedom in the system
is not sufficient  for its consistency.
Rather a separate check should be performed to ensure that the propagation of a massive higher spin field  is causal.

\setcounter{equation}0
\section{Causality Analysis for the $s-s-0$ System}\label{s-s-0causality}

Let us consider the general spin $s-s-0$ system on the base of the
same simplified model as in the previous section, where the spin 0
field is taken to be a background field.

In this case the cubic vertex will be a sum of the form

\begin{equation}\label{ss0vertex}
|V\rangle=\sum_{k=0}^{s}\frac{1}{k!((s-k)!)^2}(L^{(1)+})^{s-k}(L^{(2)+})^{s-k}(Q^{(12)+})^k c_0^1 c_0^2c_0^3|0\rangle_{123},
\end{equation}
whilst the higher spin functionals for the massive triplet can again be obtained from dimensional reduction of the massless spin-$s$ triplet.

The gauge fixing procedure for the $s-s-0$ system follows similarly as for the spin 3-3-0 system discussed in Section \ref{3-3-0gf}, so we shall omit the full details here.
Indeed, one sees again that the number of physical degrees of freedom remains unchanged when turning on the interactions provided we consider the limit of large mass $m$. In particular, we are left only with $\phi_{\mu_1\ldots\mu_s}$ which satisfies a transversality condition

\begin{equation}
\partial^{\mu_1}\phi_{\mu_1\ldots\mu_s}=0,
\end{equation}
and a nonlinear equation of motion which, in momentum space, contains the second-derivative terms
\begin{equation}\label{spinseqn}
-p^2\phi_{\mu_1\ldots\mu_s}+\sum_{k=0}^{s-2}\frac{(-2a_1^2)^{s-k}a_2^k}{k!(s-k)!}\left[p_{\mu_1}p_{\mu_2}\delta_{(\mu_{s-k+1}}^{\nu_{s-k+1}}\ldots\delta_{\mu_s)}^{\nu_s}{\tilde G}_{\mu_3\ldots\mu_{s-k}}^{\nu_1\ldots\nu_{s-k}}+\textrm{symm}\right]\phi_{\nu_1\ldots\nu_s}.
\end{equation}

Here we have already taken into account the symmetrization over indices in the $p_{\mu_i} p_{\mu_j}$ and $G_{\mu\ldots}$ terms.
For example, the relevant term for $s=4,k=0$ would look like
\begin{eqnarray}
&& \frac{2g}{3}a_1^8\left[p_{\mu_1}p_{\mu_2}{\tilde G}_{\mu_3\mu_4}^{\nu_1\nu_2\nu_3\nu_4}
+ p_{\mu_1}p_{\mu_3}{\tilde G}_{\mu_2\mu_4}^{\nu_1\nu_2\nu_3\nu_4}
+ p_{\mu_1}p_{\mu_4}{\tilde G}_{\mu_2\mu_3}^{\nu_1\nu_2\nu_3\nu_4}\right. \nonumber \\
&& \left. +p_{\mu_2}p_{\mu_3}{\tilde G}_{\mu_1\mu_4}^{\nu_1\nu_2\nu_3\nu_4}
+ p_{\mu_2}p_{\mu_4}{\tilde G}_{\mu_1\mu_3}^{\nu_1\nu_2\nu_3\nu_4}
+ p_{\mu_3}p_{\mu_4}{\tilde G}_{\mu_1\mu_2}^{\nu_1\nu_2\nu_3\nu_4}\right]\phi_{\nu_1\nu_2\nu_3\nu_4}.
\end{eqnarray}

The causality analysis of Section \ref{3-3-0causality} should then be applied to the expression \eqref{spinseqn}, i.e. we should compute the determinant of the corresponding  $\binom{\mathcal{D}+s-1}{s}\times\binom{\mathcal{D}+s-1}{s}$ matrix acting on the vector space of symmetric $s$-tensors. As before, ignoring terms of $\mathcal{O}(g^2)$, the requirement of causal propagation corresponds exactly to the case where the trace of the first-order part of this matrix vanishes.

Indeed we find

\begin{equation}\label{A-A-A}
\textrm{tr}(A)=\sum_{k=0}^{s-2}\frac{(-2a_1^2)^{s-k}\,a_2^k}{(s-k)!}\binom{s-k}{2}\binom{\mathcal{D}+s-1}{k}p_{\mu_1}p_{\mu_2}{\tilde G}_{\mu_3\ldots\mu_{s-k}}^{\mu_1\ldots\mu_{s-k}}.
\end{equation}

Using \eqref{Gtilde}
 and the zeroth-order equation for the background scalar, we find that the condition for the vanishing of $\textrm{tr}(A)$ is equivalent to finding a solution $(x,y)\in\mathbb{R}^2$ to the homogeneous equation

\begin{equation}\label{master}
\sum_{k=0}^{s-2}\frac{1}{(s-k)!}\binom{s-k}{2}\binom{\mathcal{D}+s-1}{k}x^{s-k-2}y^k=0,
\end{equation}
where $x:=2a_1^2m^2$ and $y:=a_2$.

The idea then is that any real solution $(x,y)\in\mathbb{R}^2$ to \eqref{master} corresponds to a choice of $a_1,a_2$, which
 parametrize the
 cubic vertex \eqref{ss0vertex}, such that the 
propagation of the spin $s$ degrees of freedom is within the light-cone.

We note first that, if the only real solution to \eqref{master} is $(x,y)=(0,0)$, then the theory can be causal only if it is free. Moreover, if $(x,0)$ is a solution to \eqref{master} then we must have $a_1=0$ and likewise for $a_2$.
Hence, our aim is to determine whether solutions of \eqref{master} exist with both $x$ and $y$ non-zero. Since this is the case, we can divide through by $y^{s-2}$ and reduce the problem to finding real zeroes of the degree $s-2$ polynomial

\begin{equation}\label{masterpoly}
p^{\mathcal{D}}_s(z):=\sum_{k=0}^{s-2}\frac{1}{(s-k)!}\binom{s-k}{2}\binom{\mathcal{D}+s-1}{k}z^{s-k-2},
\end{equation}
in $z=-2a_1^2m^2a_2^{-1}$.

In order to solve this problem, we first rewrite \eqref{masterpoly} as

\begin{equation}
p_s^{\mathcal{D}}(z)=\frac{1}{2}\sum_{k=0}^{s-2}\binom{\mathcal{D}+s-1}{s-2-k}\frac{z^k}{k!}=\frac{1}{2}L^{\mathcal{D}+1}_{s-2}(-z),
\end{equation}
where $L^m_n(x)$ are the generalized Laguerre polynomials \cite{Abramowitz}. The $L^m_n(x)$ are
orthogonal on $x\in [0,\infty)$ with weight function $w(x)=e^{-x}x^m$ \cite{Abramowitz} and so, from
the theory of orthogonal polynomials \cite{Szego}, have $n$ real zeroes in the range $[0,\infty)$.

Hence, we see that for any integer spin $s>2$ and in any spacetime dimension
$\mathcal{D}$, the equation $p_s^{\mathcal{D}}(z)=0$ will have $s-2$ real solutions (all with $z<0$).
Each of these zeroes (which can be found numerically if necessary) correspond to a particular codimension
1 locus in $(a_1,a_2)$ parameter space along which the cubic vertex \eqref{ss0vertex} will give rise to
causal propagation for the massive spin $s$ field coupled to a background scalar.

\section{Conclusions}
The problem of consistency of an interacting theory which contains
massive and massless higher spin fields on a flat background is
quite challenging. A necessary condition for the consistency of such a
theory  is the presence of fields with all spins up to infinity
interacting among themselves. The structure  of interaction vertices
is defined by  the requirement of the gauge  invariance, however
this condition might not be  sufficient and extra consistency
conditions, like nontriviality of the S-matrix and locality, should
be imposed \cite{Fotopoulos:2010ay}, \cite{Taronna:2011kt},
\cite{Dempster:2012vw}.

Unfortunately  a corresponding  Lagrangian theory  which contains an
infinite number of fields  and an infinite number of interacting
vertices is not known yet. Therefore, it would be useful and
instructive to try to analyse  consistency of a truncated theory
with a finite number of fields and vertices. Of course, any such 
consideration can be only approximate, however one can hope that
understanding  the structure of the truncated theory can shed some
light on properties of the general theory.

In the present paper we tried to study this problem.  We have
analyzed the theory of massive spin 3 fields coupled to a massive
scalar field using the triplet formulation for higher spin fields
\cite{Francia:2002pt}, \cite{Pashnev:1997rm}
 which is convenient for our analysis. We addressed two consistency tests:
(i) if the gauge invariance can guarantee a propagation of the true
number of physical degrees of freedom and (ii) can the gauge
invariance guarantee causal propagation? We found that the answers to
both questions are in general  negative.

First, we have considered a system of two identical massive spin $3$
and one  spin $0$ fields interacting via a cubic  vertex.  The model
is constructed in the framework of the BRST approach and is
automatically gauge invariant. However, the gauge invariance does
not fix the cubic vertex uniquely, which still contains some
free constant parameters.

After using the gauge freedom and the equations of motion, and
eliminating the auxiliary fields (including the St\"{u}ckelberg
ones) we found that the correct number of physical degrees of
freedom is not preserved by the interaction though the theory under
consideration was gauge invariant. To get the correct number of
degrees of freedom we imposed the additional restrictions on the
free parameters in the vertex.  Only after that we obtained a gauge
invariant model with the correct number of  propagating degrees of
freedom.

Second, we have analyzed the causality aspects for this model. The
only known approach to a study of the causality  in massive higher
spin field theory is the Velo-Zwanziger procedure. Unfortunately this
procedure is not directly applicable  to the model  since  it
requires the number of derivatives acting  on a dynamical field to
be at most two,  or in other words the Velo-Zwanziger procedure is
applicable for low-energy theories. Therefore to perform the
causality analysis we have derived from  the original model a
simplified low-energy  one. Finally we compared the constraints on
the parameters of the theory obtained in the two independent
procedures described above.

Let us stress once again that interacting higher spin gauge theories
require higher derivatives in the vertices. In general, the number
of these derivatives is infinite due to the infinite number of the
fields which are present in the interaction. Therefore even a negative
outcome of the Velo-Zwanziger like analysis, i.e. a possible
incompatibility of causal propagation and interactions, would have
been inconclusive for the case of the ``entire'' theory. We find
however that for a certain range of coupling constants and a mass
parameter, as well as under some conditions on a background field,
causal propagation for a spin $s$ triplet in a scalar field
background is possible. This fact apart from some interesting
implications for a low-energy theory might also have some indication
for  consistency of the ``entire" high-energy theory as well.

The main outcome of the analysis  performed in the paper is that the
gauge invariance and the presence of correct degrees of freedom for
the massive higher spin fields propagating on  a flat  background
are not enough requirements for an overall consistency. Rather one
has to perform some extra checks such as causality analysis which
can bring about extra conditions on parameters of the theory.
\vspace{5mm}

\noindent {\bf Acknowledgments.} M.T. would like to thank Department
of Physics, University of Auckland, where a part of the work has
been done, for the kind hospitality and A.Kobakhidze for useful
discussions. The work of I.L.B has been partially supported by RFBR
grant, project No. 12-02-00121-a; RFBR-DFG grant, project No.
13-02-91330; RFBR-Ukraine grant, project No. 13-02-90430; grant for
LRSS, project No 224.2012.2 and by Ministry of Education and Science
of Russian Federation, project No 14.B37.21.0774. The work of P.D.
has been supported by a STFC grant ST/1505805/1. The work of M.T.
 has been supported in part by an  Australian Research Council  grant DP120101340.
M.T. would also like to thank a grant   31/89 of Rustaveli National
Science Foundation.
\setcounter{equation}0
\appendix
\numberwithin{equation}{section}

\section{Gauge transformations. First order in $g$}\label{Appendix A}

We present here details of the contributions to the gauge transformations in Section \ref{3-3-0massive} from the terms of order $g$.

The $V_6$ part of the vertex gives the following contributions
\begin{equation}\label{order1var}
\delta_1|\Phi_i\rangle =-g\int dc_0^{(i+1)}dc_0^{(i+2)}\left[\left(\langle\Phi_{i+1}|\langle\Lambda_{i+2}|+\langle\Phi_{i+2}|\langle\Lambda_{i+1}|\right)V_6\,c_0^{(1)}c_0^{(2)}c_0^{(3)}|0\rangle_{123}\right]
\end{equation}
For the fields  in the first Hilbert space, we have

\[
\delta_1|\phi_1\rangle =\frac{g}{3!2!}\langle\phi_3|\langle\lambda_2|(L^{(1)+}_\alpha)^3(L^{(2)+}_\alpha)^2|0\rangle_{123},
\]
where we have split $L^{(i)}=L^{(i)}_\alpha+L^{(i)}_{gh}$ into a part containing oscillators and a part containing ghosts.
For component fields we find explicitly
\begin{eqnarray}
&&\delta_1 \phi_{\mu_1 \mu_2 \mu_3}\supset
\frac{ga_1^6}{2!}\left[4(\partial_{\nu_1\nu_2}\phi)(\partial_{\mu_1\mu_2\mu_3}\lambda_{\nu_1\nu_2})+4(\partial_{\nu_1}\phi)(\partial_{\mu_1\mu_2\mu_3\nu_2}\lambda_{\nu_1\nu_2})\right. \nonumber \\
&&+\phi(\partial_{\mu_1\mu_2\mu_3\nu_1\nu_2}\lambda_{\nu_1\nu_2})
 -12(\partial_{\mu_1\nu_1\nu_2}\phi)(\partial_{\mu_2\mu_3}\lambda_{\nu_1\nu_2}) -12(\partial_{\mu_1\nu_1}\phi)(\partial_{\mu_2\mu_3\nu_2}\lambda_{\nu_1\nu_2})\nonumber \\
 && -3(\partial_{\mu_1}\phi)(\partial_{\mu_2\mu_3\nu_1\nu_2}\lambda_{\nu_1\nu_2})  +12(\partial_{\mu_1\mu_2\nu_1\nu_2}\phi)(\partial_{\mu_3}\lambda_{\nu_1\nu_2})+12(\partial_{\mu_1\mu_2\nu_1}\phi)(\partial_{\mu_3\nu_2}\lambda_{\nu_1\nu_2})\nonumber \\
 && +3(\partial_{\mu_1\mu_2}\phi)(\partial_{\mu_3\nu_1\nu_2}\lambda_{\nu_1\nu_2})  -4(\partial_{\mu_1\mu_2\mu_3\nu_1\nu_2}\phi)\lambda_{\nu_1\nu_2}-4(\partial_{\mu_1\mu_2\mu_3\nu_1}\phi)(\partial_{\nu_1}\lambda_{\nu_1\nu_2}) \nonumber \\
 && \left.-(\partial_{\mu_1\mu_2\mu_3}\phi)(\partial_{\nu_1\nu_2}\lambda_{\nu_1\nu_2})\right].
\end{eqnarray}

For the scalar field in the third Hilbert space we have
\begin{equation}
\delta_1|\phi_3\rangle =-\frac{g}{3!}\langle\phi_1|\langle\lambda_2|(L_\alpha^{(1)+})^3(L_\alpha^{(2)+})^2|0\rangle_{123}
+\frac{g}{2!}\langle C_1|\langle\lambda_2|(L_\alpha^{(1)+})^2(L_\alpha^{(2)+})^2|0\rangle_{123},
\end{equation}
from which we find
\begin{eqnarray}
&& \delta_1\phi \supset -\frac{ga_1^6}{3!}\left[32(\partial_{\nu_1\nu_2}\phi_{\mu_1\mu_2\mu_3})(\partial_{\mu_1\mu_2\mu_3}\lambda_{\nu_1\nu_2})+32(\partial_{\nu_1}\phi_{\mu_1\mu_2\mu_3})(\partial_{\mu_1\mu_2\mu_3\nu_2}\lambda_{\nu_1\nu_2})
\right. \nonumber \\
&& +8\phi_{\mu_1\mu_2\mu_3}(\partial_{\mu_1\mu_2\mu_3\nu_1\nu_2}\lambda_{\nu_1\nu_2})+48(\partial_{\mu_1\nu_1\nu_2}\phi_{\mu_1\mu_2\mu_3})(\partial_{\mu_2\mu_3}\lambda_{\nu_1\nu_2}) \nonumber \\
&& +48(\partial_{\mu_1\nu_1}\phi_{\mu_1\mu_2\mu_3})(\partial_{\mu_2\mu_3\nu_2}\lambda_{\nu_1\nu_2})  +12(\partial_{\mu_1}\phi_{\mu_1\mu_2\mu_3})(\partial_{\mu_2\mu_3\nu_1\nu_2}\lambda_{\nu_1\nu_2}) \nonumber \\
&& +24(\partial_{\mu_1\mu_2\nu_1\nu_2}\phi_{\mu_1\mu_2\mu_3})(\partial_{\mu_3}\lambda_{\nu_1\nu_2})  +24(\partial_{\mu_1\mu_2\nu_1}\phi_{\mu_1\mu_2\mu_3})(\partial_{\mu_3\nu_2}\lambda_{\nu_1\nu_2}) \nonumber \\
&& +6(\partial_{\mu_1\mu_2}\phi_{\mu_1\mu_2\mu_3})(\partial_{\mu_3\nu_1\nu_2}\lambda_{\nu_1\nu_2}) +4(\partial_{\mu_1\mu_2\mu_3\nu_1\nu_2}\phi_{\mu_1\mu_2\mu_3})\lambda_{\nu_1\nu_2} \nonumber \\
&& \left. +4(\partial_{\mu_1\mu_2\mu_3\nu_1}\phi_{\mu_1\mu_2\mu_3})(\partial_{\nu_2}\lambda_{\nu_1\nu_2})+(\partial_{\mu_1\mu_2\mu_3}\phi_{\mu_1\mu_2\mu_3})(\partial_{\nu_1\nu_2}\lambda_{\nu_1\nu_2})\right] \nonumber \\
&& + \frac{ga_1^6}{2!}\left[16(\partial_{\nu_1\nu_2}C_{\mu_1\mu_2})(\partial_{\mu_1\mu_2}\lambda_{\nu_1\nu_2})+16(\partial_{\nu_1}C_{\mu_1\mu_2})(\partial_{\mu_1\mu_2\nu_2}\lambda_{\nu_1\nu_2})
\right. \nonumber \\
&& +4C_{\mu_1\mu_2}(\partial_{\mu_1\mu_2\nu_1\nu_2}\lambda_{\nu_1\nu_2})+16(\partial_{\mu_1\nu_1\nu_2}C_{\mu_1\mu_2})(\partial_{\mu_2}\lambda_{\nu_1\nu_2})+16(\partial_{\mu_1\nu_1}C_{\mu_1\mu_2})(\partial_{\mu_2\nu_2}\lambda_{\nu_1\nu_2}) \nonumber \\
&& +4(\partial_{\mu_1}C_{\mu_1\mu_2})(\partial_{\mu_2\nu_1\nu_2}\lambda_{\nu_1\nu_2}) +4(\partial_{\mu_1\mu_2\nu_1\nu_2}C_{\mu_1\mu_2})\lambda_{\nu_1\nu_2} +4(\partial_{\mu_1\mu_2\nu_1}C_{\mu_1\mu_2})(\partial_{\nu_2}\lambda_{\nu_1\nu_2}) \nonumber \\
&& \left. +(\partial_{\mu_1\mu_2}C_{\mu_1\mu_2})(\partial_{\nu_1\nu_2}\lambda_{\nu_1\nu_2})\right]. \nonumber \\
\end{eqnarray}
In a similar way  we find that for $V_6$  the linear gauge transformations of $|C\rangle$ and $|D\rangle$ are unmodified.

The vertex $V_5$ gives the following contribution to the gauge transformations
\begin{equation}\label{v5tr}
\delta_1|\Phi_i\rangle =-g\int dc_0^{(i+1)}dc_0^{(i+2)}\left[\left(\langle\Phi_{i+1}|\langle\Lambda_{i+2}|+\langle\Phi_{i+2}|\langle\Lambda_{i+1}|\right)V_5\,c_0^{(1)}c_0^{(2)}c_0^{(3)}|0\rangle_{123}\right]
\end{equation}
If we introduce for convenience the notation
\begin{equation}\label{qhatdef}
Q^{(12)}=\hat{Q}^{(12)}+\frac{a_2}{2a_1m_1}\alpha^{(1)}_DL^{(2)} -\frac{a_2}{2a_1m_2}\alpha^{(2)}_DL^{(1)}, \hspace{5mm} \hat{Q}^{(12)}=\hat{Q}^{(12)}_\alpha +\hat{Q}^{(12)}_{gh.},
\end{equation}
then the transformations (\ref{v5tr})
 translate into
\begin{eqnarray}
\delta_1|\phi_1\rangle &=& g\langle\phi_3|\langle\lambda_2|\left[\frac{1}{2}(L^{(1)+}_\alpha)^2 L^{(2)+}_\alpha \hat{Q}^{(12)+}_\alpha +\frac{3}{8m}\alpha^{(1)+}_D(L^{(1)+}_\alpha)^2(L^{(2)+}_\alpha)^2\right. \nonumber \\
&& \left. \hspace{18mm}-\frac{1}{4m}\alpha^{{(2)}+}_D(L^{(1)+}_\alpha)^3 L^{(2)+}_\alpha\right]|0\rangle_{123},
\end{eqnarray}
\begin{equation}
\delta_1|C_1\rangle =\frac{g}{8}\langle\phi_3|\langle\lambda_2|(L^{(1)+}_\alpha)^2(L^{(2)+}_\alpha)^2|0\rangle_{123},
\end{equation}
and

\begin{eqnarray}
&& \delta_1|\phi_3\rangle =-g\langle\phi_1|\langle\lambda_2|(L^{(1)+}_\alpha)^2(L^{(2)+}_\alpha)^2\hat{Q}_\alpha^{(12)+}|0\rangle_{123}  \nonumber \\
&& +2g\langle C_1|\langle\lambda_2|L^{(1)+}_\alpha L^{(2)+}_\alpha\hat{Q}_\alpha^{(12)+}|0\rangle_{123}
-\frac{g}{2}\langle D_1|\langle\lambda_2|L^{(1)+}_\alpha(L^{(2)+}_\alpha)^2|0\rangle_{123}
 \nonumber \\
&& -\frac{3g}{4m}\langle\phi_1|\langle\lambda_2|\alpha^{(1)+}_D(L^{(1)+}_\alpha)^2(L^{(2)+}_\alpha)^2|0\rangle_{123}
+\frac{3g}{2m}\langle C_1|\langle\lambda_2|\alpha^{(1)+}_D L^{(1)+}_\alpha(L^{(2)+}_\alpha)^2|0\rangle_{123}
 \nonumber \\
&& +\frac{g}{2m}\langle\phi_1|\langle\lambda_2|\alpha^{(2)+}_D(L^{(1)+}_\alpha)^3 L^{(2)+}_\alpha|0\rangle_{123}
-\frac{3g}{2m}\langle C_1|\langle\lambda_2|\alpha^{(2)+}_D (L^{(1)+}_\alpha)^2 L^{(2)+}_\alpha|0\rangle_{123}, \nonumber \\
\end{eqnarray}
whereas the variation for $|D_1 \rangle$ is not modified.
Equivalently one has
\begin{eqnarray}
&&\delta_1\phi_{\mu_1\mu_2\mu_3}\supset 3ga_1^4a_2\left[2(\partial_{\nu_1}\phi)(\partial_{\mu_1\mu_2}\lambda_{\nu_1\mu_3})+\phi(\partial_{\nu_1\mu_1\mu_2}\lambda_{\nu_1\mu_3})-4(\partial_{\mu_1\nu_1}\phi)(\partial_{\mu_2}\lambda_{\nu_1\mu_3})
\right. \nonumber \\
&& \left.-2(\partial_{\mu_1}\phi)(\partial_{\mu_2\nu_1}\lambda_{\nu_1\mu_3})+2(\partial_{\mu_1\mu_2\nu_1}\phi)\lambda_{\nu_1\mu_3}+(\partial_{\mu_1\mu_2}\phi)(\partial_{\nu_1}\lambda_{\nu_1\mu_3})\right] \nonumber \\
&& -\frac{3ga_1^4a_2}{2m}\left[2(\partial_{\nu_1}\phi)(\partial_{\mu_1\mu_2\mu_3}\lambda_{\nu_1})+\phi(\partial_{\mu_1\mu_2\mu_3\nu_1}\lambda_{\nu_1})-6(\partial_{\mu_1\nu_1}\phi)(\partial_{\mu_2\mu_3}\lambda_{\nu_1})
\right. \nonumber \\
&& -3(\partial_{\mu_1}\phi)(\partial_{\mu_2\mu_3\nu_1}\lambda_{\nu_1})+6(\partial_{\mu_1\mu_2\nu_1}\phi)(\partial_{\mu_3}\lambda_{\nu_1})+3(\partial_{\mu_1\mu_2}\phi)(\partial_{\mu_3\nu_1}\lambda_{\nu_1}) \nonumber \\
&& \left. -2(\partial_{\mu_1\mu_2\mu_3\nu_1}\phi)\lambda_{\nu_1}-(\partial_{\mu_1\mu_2\mu_3}\phi)(\partial_{\nu_1}\lambda_{\nu_1})\right],
\end{eqnarray}

\begin{eqnarray}
&& \delta_1 h_{\mu_1\mu_2}\supset
 -\frac{ga_1^4a_2}{2}\left[2(\partial_{\nu_1}\phi)(\partial_{\mu_1\mu_2}\lambda_{\nu_1})+\phi(\partial_{\mu_1\mu_2\nu_1}\lambda_{\nu_1})-4(\partial_{\mu_1\nu_1}\phi)(\partial_{\mu_2}\lambda_{\nu_1})
\right. \nonumber \\
&&  -2 \left(\partial_{\mu_1}\phi)(\partial_{\mu_2\nu_1}\lambda_{\nu_1})+2(\partial_{\mu_1\mu_2\nu_1}\phi)\lambda_{\nu_1}+(\partial_{\mu_1\mu_2}\phi)(\partial_{\nu_1}\lambda_{\nu_1})\right] \nonumber \\
&& -\frac{3ga_1^4a_2}{4m}\left[4(\partial_{\nu_1\nu_2}\phi)(\partial_{\mu_1\mu_2}\lambda_{\nu_1\nu_2})+4(\partial_{\nu_1}\phi)(\partial_{\mu_1\mu_2\nu_2}\lambda_{\nu_1\nu_2})+\phi(\partial_{\mu_1\mu_2\nu_1\nu_2}\lambda_{\nu_1\nu_2})
\right. \nonumber \\
&& -8(\partial_{\mu_1\nu_1\nu_2}\phi)(\partial_{\mu_2}\lambda_{\nu_1\nu_2})-8(\partial_{\mu_1\nu_1}\phi)(\partial_{\mu_2\nu_2}\lambda_{\nu_1\nu_2})-2(\partial_{\mu_1}\phi)(\partial_{\mu_2\nu_1\nu_2}\lambda_{\nu_1\nu_2}) \nonumber \\
&&  +4 \left(\partial_{\mu_1\mu_2\nu_1\nu_2}\phi)\lambda_{\nu_1\nu_2}+4(\partial_{\mu_1\mu_2\nu_1}\phi)(\partial_{\nu_2}\lambda_{\nu_1\nu_2})+(\partial_{\mu_1\mu_2}\phi)(\partial_{\nu_1\nu_2}\lambda_{\nu_1\nu_2})\right] ,
\end{eqnarray}

\begin{eqnarray}
&& \delta_1 C_{\mu_1\mu_2} \supset \frac{ga_1^4a_2}{4}\left[4(\partial_{\nu_1\nu_2}\phi)(\partial_{\mu_1\mu_2}\lambda_{\nu_1\nu_2})+4(\partial_{\nu_1}\phi)(\partial_{\mu_1\mu_2\nu_2}\lambda_{\nu_1\nu_2})
\right. \nonumber \\
&& +\phi(\partial_{\mu_1\mu_2\nu_1\nu_2}\lambda_{\nu_1\nu_2})-8(\partial_{\mu_1\nu_1\nu_2}\phi)(\partial_{\mu_2}\lambda_{\nu_1\nu_2})-8(\partial_{\mu_1\nu_1}\phi)(\partial_{\mu_2\nu_2}\lambda_{\nu_1\nu_2}) \nonumber \\
&&  -2(\partial_{\mu_1}\phi)(\partial_{\mu_2\nu_1\nu_2}\lambda_{\nu_1\nu_2})+4(\partial_{\mu_1\mu_2\nu_1\nu_2}\phi)\lambda_{\nu_1\nu_2}+4(\partial_{\mu_1\mu_2\nu_1}\phi)(\partial_{\nu_2}\lambda_{\nu_1\nu_2}) \nonumber \\
&& \left. +(\partial_{\mu_1\mu_2}\phi)(\partial_{\nu_1\nu_2}\lambda_{\nu_1\nu_2})\right] ,
\end{eqnarray}
and

\begin{eqnarray}
&& \delta_1 \phi \supset -ga_1^4a_2\left[8(\partial_{\nu_1}\phi_{\mu_1\mu_2\mu_3})(\partial_{\mu_1\mu_2}\lambda_{\nu_1\mu_3})
+4\phi_{\mu_1\mu_2\mu_3}(\partial_{\mu_1\mu_2\nu_1}\lambda_{\nu_1\mu_3}) \right. \nonumber \\
&&
+8(\partial_{\mu_1\nu_1}\phi_{\mu_1\mu_2\mu_3})(\partial_{\mu_2}\lambda_{\nu_1\mu_3})
+4(\partial_{\mu_1}\phi_{\mu_1\mu_2\mu_3})(\partial_{\mu_2\nu_1}\lambda_{\nu_1\mu_3}) \nonumber \\
&& \left. +2(\partial_{\mu_1\mu_2\nu_1}\phi_{\mu_1\mu_2\mu_3})\lambda_{\nu_1\mu_3}
+(\partial_{\mu_1\mu_2}\phi_{\mu_1\mu_2\mu_3})(\partial_{\nu_1}\lambda_{\nu_1\mu_3})\right] \nonumber \\
&& +\frac{ga_1^4a_2}{2}\left[8(\partial_{\nu_1}h_{\mu_1\mu_2})(\partial_{\mu_1\mu_2}\lambda_{\nu_1})
+4h_{\mu_1\mu_2}(\partial_{\mu_1\mu_2\nu_1}\lambda_{\nu_1})
+8(\partial_{\mu_1\nu_1}h_{\mu_1\mu_2})(\partial_{\mu_2}\lambda_{\nu_1})\right. \nonumber \\
&&
\left. +4(\partial_{\mu_1}h_{\mu_1\mu_2})(\partial_{\mu_2\nu_1}\lambda_{\nu_1}) +2(\partial_{\mu_1\mu_2\nu_1}h_{\mu_1\mu_2})\lambda_{\nu_1}
+(\partial_{\mu_1\mu_2}h_{\mu_1\mu_2})(\partial_{\nu_1}\lambda_{\nu_1})\right] \nonumber \\
&& +2ga_1^4a_2\left[4(\partial_{\nu_1}C_{\mu_1\mu_2})(\partial_{\mu_1}\lambda_{\nu_1\mu_2})
+2C_{\mu_1\mu_2}(\partial_{\mu_1\nu_1}\lambda_{\nu_1\mu_2})
\right. \nonumber \\
&& \left. +2(\partial_{\mu_1\nu_1}C_{\mu_1\mu_2})\lambda_{\nu_1\mu_2}
+(\partial_{\mu_1}C_{\mu_1\mu_2})(\partial_{\nu_1}\lambda_{\nu_1\mu_2})\right] \nonumber \\
&& +ga_1^4a_2\left[4(\partial_{\nu_1}C_{\mu_1})(\partial_{\mu_1}\lambda_{\nu_1})
+2C_{\mu_1}(\partial_{\mu_1\nu_1}\lambda_{\nu_1})
 +2(\partial_{\mu_1\nu_1}C_{\mu_1})\lambda_{\nu_1}
+(\partial_{\mu_1}C_{\mu_1})(\partial_{\nu_1}\lambda_{\nu_1})\right] \nonumber \\
&& +\frac{ga_1^4a_2}{2}\left[8(\partial_{\nu_1\nu_2}D_{\mu_1})(\partial_{\mu_1}\lambda_{\nu_1\nu_2})+8(\partial_{\nu_1}D_{\mu_1})(\partial_{\mu_1\nu_2}\lambda_{\nu_1\nu_2})
+2D_{\mu_1}(\partial_{\mu_1\nu_1\nu_2}\lambda_{\nu_1\nu_2})
\right. \nonumber \\
&& \left. +4(\partial_{\mu_1\nu_1\nu_2}D_{\mu_1})\lambda_{\nu_1\nu_2}+4(\partial_{\mu_1\nu_1}D_{\mu_1})(\partial_{\nu_2}\lambda_{\nu_1\nu_2})+(\partial_{\mu_1}D_{\mu_1})(\partial_{\nu_1\nu_2}\lambda_{\nu_1\nu_2})\right] \nonumber \\
&& +\frac{3ga_1^4a_2}{4m}\left[16(\partial_{\nu_1\nu_2}h_{\mu_1\mu_2})(\partial_{\mu_1\mu_2}\lambda_{\nu_1\nu_2})+16(\partial_{\nu_1}h_{\mu_1\mu_2})(\partial_{\mu_1\mu_2\nu_2}\lambda_{\nu_1\nu_2})
\right. \nonumber \\
&& +4h_{\mu_1\mu_2}(\partial_{\mu_1\mu_2\nu_1\nu_2}\lambda_{\nu_1\nu_2})
+16(\partial_{\mu_1\nu_1\nu_2}h_{\mu_1\mu_2})(\partial_{\mu_2}\lambda_{\nu_1\nu_2}) +16(\partial_{\mu_1\nu_1}h_{\mu_1\mu_2})(\partial_{\mu_2\nu_2}\lambda_{\nu_1\nu_2}) \nonumber \\
&& +4(\partial_{\mu_1}h_{\mu_1\mu_2})(\partial_{\mu_2\nu_1\nu_2}\lambda_{\nu_1\nu_2})
+4(\partial_{\mu_1\mu_2\nu_1\nu_2}h_{\mu_1\mu_2})\lambda_{\nu_1\nu_2}
+4(\partial_{\mu_1\mu_2\nu_1}h_{\mu_1\mu_2})(\partial_{\nu_2}\lambda_{\nu_1\nu_2}) \nonumber \\
&& \left. +(\partial_{\mu_1\mu_2}h_{\mu_1\mu_2})(\partial_{\nu_1\nu_2}\lambda_{\nu_1\nu_2})\right] \nonumber \\
&& +\frac{3ga_1^4a_2}{2m}\left[8(\partial_{\nu_1\nu_2}C_{\mu_1})(\partial_{\mu_1}\lambda_{\nu_1\nu_2})
+8(\partial_{\nu_1}C_{\mu_1})(\partial_{\mu_1\nu_2}\lambda_{\nu_1\nu_2})
+2C_{\mu_1}(\partial_{\mu_1\nu_1\nu_2}\lambda_{\nu_1\nu_2})
\right. \nonumber \\
&&
\left. +4(\partial_{\mu_1\nu_1\nu_2}C_{\mu_1})\lambda_{\nu_1\nu_2}
+4(\partial_{\mu_1\nu_1}C_{\mu_1})(\partial_{\nu_2}\lambda_{\nu_1\nu_2})
+(\partial_{\mu_1}C_{\mu_1})(\partial_{\nu_1\nu_2}\lambda_{\nu_1\nu_2})\right] \nonumber \\
&& -\frac{ga_1^4a_2}{2m}\left[16(\partial_{\nu_1}\phi_{\mu_1\mu_2\mu_3})(\partial_{\mu_1\mu_2\mu_3}\lambda_{\nu_1})+8\phi_{\mu_1\mu_2\mu_3}(\partial_{\mu_1\mu_2\mu_3\nu_1}\lambda_{\nu_1})
\right. \nonumber \\
&& +24(\partial_{\mu_1\nu_1}\phi_{\mu_1\mu_2\mu_3})(\partial_{\mu_2\mu_3}\lambda_{\nu_1})
+12(\partial_{\mu_1}\phi_{\mu_1\mu_2\mu_3})(\partial_{\mu_2\mu_3\nu_1}\lambda_{\nu_1}) \nonumber \\
&& +12(\partial_{\mu_1\mu_2\nu_1}\phi_{\mu_1\mu_2\mu_3})(\partial_{\mu_3}\lambda_{\nu_1})
+6(\partial_{\mu_1\mu_2}\phi_{\mu_1\mu_2\mu_3})(\partial_{\mu_3\nu_1}\lambda_{\nu_1}) \nonumber \\
&& \left. +2(\partial_{\mu_1\mu_2\mu_3\nu_1}\phi_{\mu_1\mu_2\mu_3})\lambda_{\nu_1}
+(\partial_{\mu_1\mu_2\mu_3}\phi_{\mu_1\mu_2\mu_3})(\partial_{\nu_1}\lambda_{\nu_1})\right] \nonumber \\
&& +\frac{3ga_1^4a_2}{2m}\left[8(\partial_{\nu_1}C_{\mu_1\mu_2})(\partial_{\mu_1\mu_2}\lambda_{\nu_1}) +4C_{\mu_1\mu_2}(\partial_{\mu_1\mu_2\nu_1}\lambda_{\nu_1}
+8(\partial_{\mu_1\nu_1}C_{\mu_1\mu_2})(\partial_{\mu_2}\lambda_{\nu_1})
\right. \nonumber \\
&& \left. +4(\partial_{\mu_1}C_{\mu_1\mu_2})(\partial_{\mu_2\nu_1}\lambda_{\nu_1})
+2(\partial_{\mu_1\mu_2\nu_1}C_{\mu_1\mu_2})\lambda_{\nu_1}
+(\partial_{\mu_1\mu_2}C_{\mu_1\mu_2})(\partial_{\nu_1}\lambda_{\nu_1})\right].
\end{eqnarray}

\section{Examples of the lower spin fields}\label{Appendix B}

\subsection{$2-2-0$ Vertex}

We start with the massive spin two ``triplets''
\begin{eqnarray}
|\Phi_{1,2}\rangle &=& 
\frac{1}{2!}h_{\mu_1\mu_2}(x)\alpha^{(1,2)+}_{\mu_1}\alpha^{(1,2)+}_{\mu_2}|0\rangle_{1,2}
+ib_{\mu_1}(x)\alpha^{(1,2)+}_{\mu_1}\alpha^{(1,2)+}_D|0\rangle_{1,2} \nonumber \\
&& +\varphi(x)\alpha^{(1,2)+}_D\alpha^{(1,2)+}_D|0\rangle_{1,2}-iC_{\mu_1}(x)\alpha^{(1,2)+}_{\mu_1}c_0^{(1,2)}b^{(1,2)+}|0\rangle_{1,2} \nonumber \\
&& -C(x)\alpha^{(1,2)+}_D c_0^{(1,2)}b^{(1,2)+}|0\rangle_{1,2} +D(x)c^{(1,2)+}b^{(1,2)+}|0\rangle_{1,2}.
\end{eqnarray}
Here the fields $b_\mu$ and $\varphi$ are   St\"{u}ckelberg  fields. 

The  cubic interaction vertices are given in terms of (\ref{f-1})--(\ref{f-2}) as
\begin{equation}
V_4=\frac{1}{2!2!}(L^{(1)+})^2(L^{(2)+})^2, \hspace{3mm}
 V_3=L^{(1)+}L^{(2)+}Q^{(12)+}, \hspace{3mm} 
 V_2=\frac{1}{2!}(Q^{(12)+})^2.
\end{equation}

Provided one  can consistently impose the transversality condition $\partial^{\mu_1}h_{\mu_1\mu_2}=0$, 
one can see from (\ref{A-A-A})
  that the requirement of causal propagation is equivalent to the condition
\begin{equation}
\frac{(-2a_1^2)^2}{2!}(-m^2)=0,
\end{equation}
which is only satisfied for $a_1=0$, and hence the vertices $V_4$ and $V_3$ vanish.

The equations of motion coming from the Lagrangian with the remaining vertex $V_2$ are as follows.

With respect to  $h_{\mu_1\mu_2}$ 
\begin{eqnarray}
&& (\Box -m^2)h_{\mu_1\mu_2}=\partial_{(\mu_1}C_{\mu_2)}-ga_2^2 h_{\mu_1\mu_2}\phi \nonumber \\
&& +\frac{ga_2^2}{m}\left[(\partial_{\mu_1}b_{\mu_2})\phi -b_{\mu_2}(\partial_{\mu_1}\phi)\right] \nonumber \\
&& +\frac{ga_2^2}{m}\left[(\partial_{\mu_1\mu_2}\varphi)\phi -2(\partial_{\mu_1}\varphi)(\partial_{\mu_2}\phi)+\varphi(\partial_{\mu_1\mu_2}\phi)\right].
\end{eqnarray}

With respect to  $b_{\mu_1}$ 
\begin{eqnarray}
&& (\Box -m^2)b_{\mu_1} =-\partial_{\mu_1}C+mC_{\mu_1} +\frac{ga_2^2}{2!}b_{\mu_1}\phi \nonumber \\
&& -\frac{ga_2^2}{2m}\left[2h_{\mu_1\nu_1}(\partial_{\nu_1}\phi)+(\partial_{\nu_1}h_{\mu_1\nu_1})\phi\right] \nonumber \\
&& -\frac{ga_2^2}{2m}\left[(\partial_{\mu_1}\varphi)\phi -\varphi(\partial_{\mu_1}\phi)\right] \\
&& +\frac{ga_2^2}{4m^2}\left[2(\partial_{\mu_1}b_{\nu_1})(\partial_{\nu_1}\phi)-2b_{\nu_1}(\partial_{\mu_1\nu_1}\phi)
+(\partial_{\mu_1\nu_1}b_{\nu_1})\phi -(\partial_{\nu_1}b_{\nu_1})(\partial_{\mu_1}\phi)\right]. \nonumber
\end{eqnarray}

With respect to   $\varphi$
\begin{eqnarray}
&& (\Box -m^2)\varphi =mC-\frac{ga_2^2}{4}\varphi\phi \nonumber \\
&& +\frac{ga_2^2}{4m}\left[2b_{\nu_1}(\partial_{\nu_1}\phi)+(\partial_{\nu_1}b_{\nu_1})\phi\right] \nonumber \\
&& +\frac{ga_2^2}{4m^2}\left[4h_{\nu_1\nu_2}(\partial_{\nu_1\nu_2}\phi)+4(\partial_{\nu_1}h_{\nu_1\nu_2})(\partial_{\nu_2}\phi)+(\partial_{\nu_1\nu_2}h_{\nu_1\nu_2})\phi\right].
\end{eqnarray}

With respect to $C_{\mu_1}$
\begin{equation}
C_{\mu_1}=\partial_{\mu_2}h_{\mu_1\mu_2}-mb_{\mu_1}-\partial_{\mu_1}D.
\end{equation}

With respect to  $C$ 
\begin{equation}
C=-\partial_{\mu_1}b_{\mu_1} -2m\varphi +mD +\frac{ga_2^2}{4m}D\phi +\frac{ga_2^2}{4m}C\phi.
\end{equation}

And finally with respect to  $D$ 
\begin{equation}
(\Box -m^2)D=\partial_{\mu_1}C_{\mu_1}+mC-\frac{g}{2}D\phi -\frac{ga_2^2}{4m}C\phi.
\end{equation}

At zeroth-order, the gauge transformations are given by 
\begin{eqnarray}
\delta_0 h_{\mu_1\mu_2} &=& \partial_{(\mu_1}\lambda_{\mu_2)}, \nonumber \\
\delta_0 b_{\mu_1} &=& -\partial_{\mu_1}\lambda +m\lambda_{\mu_1}, \nonumber \\
\delta_0 \varphi &=& m\lambda, \\
\delta_0 C_{\mu_1} &=& (\Box -m^2)\lambda_{\mu_1}, \nonumber \\
\delta_0 C &=& (\Box -m^2)\lambda, \nonumber \\
\delta_0 D &=& \partial_{\mu_1}\lambda_{\mu_1}+m\lambda. \nonumber
\end{eqnarray}

As with the $3-3-0$ case, we can use the gauge parameters $\lambda$ and $\lambda_\mu$ to gauge away the fields $b_\mu$ and $\varphi$.

In order that the interacting system describes the correct number of degrees of freedom for a massive spin 2 field, we must be able to consistently impose the transversality constraint $\partial^{\mu}h_{\mu\nu}=0$ along with setting the auxiliary fields $C,C_{\mu_1},D$ to zero. 
It can be easily achieved by taking a constant background field  $\phi=\langle\phi\rangle$. The wave equation for the massive spin two field becomes 
 \begin{equation}
(\Box -m^2)h_{\mu_1\mu_2}=-ga_2^2 h_{\mu_1\mu_2}\phi,
\end{equation}
from which it is obvious that one has just a redefinition of the constant mass parameter.

\subsection{$1-1-0$ Vertex}

Again we start with the massive spin-1 ``triplets''
\begin{eqnarray}
|\Phi_{1,2}\rangle &=& 
+A_{\mu_1}(x)\alpha^{(1,2)+}_{\mu_1}|0\rangle_{1,2}
+i\varphi(x)\alpha^{(1,2)+}_D|0\rangle_{1,2} \nonumber \\
&& -iC(x)c_0^{(1,2)}b^{(1,2)+}|0\rangle_{1,2}.
\end{eqnarray}
Here the field $\varphi$ is  a  St\"{u}ckelberg field. 

The cubic  interaction vertices  are given in terms of (\ref{f-1})--(\ref{f-2}) as
\begin{equation}
V_2=L^{(1)+}L^{(2)+}, \hspace{3mm}
 V_1=Q^{(12)+}.
\end{equation}
After imposing the transversality condition $\partial^{\mu}A_{\mu}=0$, neither of these vertices will give rise to terms in the equation of motion for $A_{\mu}$ with two derivatives. Hence, the requirement of causality adds no additional constraints on the parameters $a_1,a_2$.

The equations of motion resulting from the Lagrangian with vertex $V_2+V_1$ are as follows.

With respect to  $A_{\mu_1}$
\begin{eqnarray}
&& (\Box -m^2)A_{\mu_1} = \partial_{\mu_1}C \nonumber \\
&& +ga_1^2\left[2(\partial_{\mu_1}A_{\nu_1})(\partial_{\nu_1}\phi)+(\partial_{\mu_1\nu_1}A_{\nu_1})\phi -2A_{\nu_1}(\partial_{\mu_1\nu_1}\phi)-(\partial_{\nu_1}A_{\nu_1})(\partial_{\mu_1}\phi)\right] \nonumber \\
&& -ga_2 A_{\mu_1}\phi 
 +\frac{ga_2}{2m}\left[(\partial_{\mu_1}\varphi)\phi-\varphi(\partial_{\mu_1}\phi)\right].
\end{eqnarray}
With respect to  $\varphi$ 
\begin{equation}
(\Box -m^2)\varphi =mC-ga_2\varphi\phi 
-\frac{ga_2}{2m}\left[2A_{\nu_1}(\partial_{\nu_1}\phi)+(\partial_{\nu_1}A_{\nu_1})\phi\right].
\end{equation}
And finally with respect to  $C$ 
\begin{equation}
C=\partial_{\mu_1}A_{\mu_1}-m\varphi -ga_1^2 C\phi.
\end{equation}
At zeroth-order, the gauge transformations are given by 
\begin{eqnarray}
\delta_0 A_{\mu_1} &=& \partial_{\mu_1}\lambda, \nonumber \\
\delta_0 \varphi &=& m\lambda, \\
\delta_0 C &=& (\Box -m^2)\lambda. \nonumber
\end{eqnarray}
We can use the gauge parameter $\lambda$ to gauge away the field $\varphi$. Further,
in order that the interacting system describes the correct number of degrees of freedom for a massive spin one field, one must be able to consistently impose the transversality constraint $\partial^{\mu}A_{\mu}=0$ along with setting the field $C$ to zero.
Again one can easily check that this can be satisfied for a constant background scalar field  $\phi=\langle\phi\rangle$.
The wave equation for the spin one field  
$A_{\mu}$ is
\begin{equation}
(\Box -m^2)A_{\mu}=-ga_2 A_{\mu_1}\langle\phi\rangle,
\end{equation}
 and therefore one has a simple redefinition of mass parameter, as it was for the case of the massive spin two field.

\end{document}